\documentclass[%
 reprint,
 amsmath,
 amssymb,
 aps,
 pra,
]{revtex4-2}

\usepackage{graphicx}
\usepackage{dcolumn}
\usepackage{bm}
\usepackage{quantikz}
\usepackage[all]{xy} 
\usepackage{enumerate}
\usepackage{amsmath,amssymb}
\usepackage{mathrsfs}
\usepackage[justification=justified]{caption}
\usepackage{xcolor}
\usepackage{verbatim}
\usepackage[normalem]{ulem}
\usepackage[rightcaption]{sidecap}
\usepackage{graphicx}
\usepackage{subcaption}
\usepackage{caption}
\usepackage{multirow}
\usepackage{amsthm,epsfig}
\usepackage{pgfplots}
\pgfplotsset{compat=1.14}
\usepackage{float}
\usepackage{multirow}
\usepackage[colorlinks,citecolor=black,urlcolor=black,linkcolor =black,bookmarks=false,hypertexnames=true]{hyperref} 
\usepackage{relsize}
\usepackage{forest,adjustbox}
\usepackage{dsfont}
\usepackage{amscd}
\usepackage{boldline}
\usepackage{tabularx}
\usepackage{stackengine}

\theoremstyle{plain}

\newtheorem{proposition}[equation]{Proposition}

\newtheorem{claim}[equation]{Claim}
\theoremstyle{definition}
\newtheorem{definition}[equation]{Definition}
\newtheorem{remark}[equation]{Remark}
\newtheorem{hypothesis}[equation]{Hypothesis}

\newcommand{\Tr}{\operatorname{Tr}}
\DeclareMathOperator{\SU}{\mbox{SU}}
\DeclareMathOperator{\fe}{\mbox{Feat}}
\DeclareMathOperator{\manc}{\mbox{EmbedMan}^{ \mathbb{C} }}
\DeclareMathOperator{\man}{\mbox{EmbedMan}}
\DeclareMathOperator{\Dist}{\mathrm{d}}

\DeclareFontFamily{OT1}{pzc}{}
\DeclareFontShape{OT1}{pzc}{m}{it}{ <-> s*[1.2] pzcmi7t }{}
\DeclareMathAlphabet{\mathpzc}{OT1}{pzc}{m}{it}

\def\R{{{\mathbb R}}}
\def\C{{{\mathbb C}}}

\def\dim{{{\mbox dim}}}

\newcommand\define[1]{\emph{\textbf{#1}}}

\begin{document}

\title{Quantum Circuits, Feature Maps, and Expanded Pseudo-Entropy: Analysis of Encoding Real-World Data into a Quantum Computer}

\author{Andrew Vlasic}
\email{avlasic@deloitte.com}
\affiliation{
Deloitte Consulting LLP
}


\author{Payal Solanki}
\affiliation{
Deloitte Consulting LLP
}

\author{Anh Pham}
\affiliation{
Deloitte Consulting LLP
}

\date{\today}

\begin{abstract}
This manuscript introduces a computationally efficient method to calculate the nonlinearity of a quantum feature map, as well as a method for determining whether a quantum feature map will have a high concentration of quantum states. The technique analyzes quantum operators, through an extension of the functions of von Neumann entropy and state-transition pseudo-entropy, by deriving a method to extract the entropy of an operator. The technique is denoted as \textit{operator pseudo-entropy}, is rigorously derived, and is generally complex valued; as with state-transition pseudo-entropy, complex values contain a lot of information about entanglement or nonlinearity. The characteristics of a class of quantum feature maps are rigorously shown. The operator pseudo-entropy is illuminated through experiments and compared with von Neumann entropy and state-transition pseudo-entropy. We end the manuscript with open questions and potential directions for further research.

\end{abstract}


\maketitle

\section{\label{sec:intro}Introduction}
We introduce an analytic extension of von Neumann entropy in which the domain is the space of special unitary operators, denoted as \define{operator pseudo-entropy}, or \textbf{pseudo-entropy}, for short. This formulation of pseudo-entropy yields a technique to quickly and efficiently calculate the intensity of nonlinearity of a quantum feature map \cite{havlivcek2019supervised,schuld2019quantum,schuld2021effect}, as well as the concentration of quantum state generated \cite{thanasilp2024exponential}. Particularly, operator pseudo-entropy calculates the effect of entanglement in the system without the need to calculate quantum states. The effect is captured as complex values, and interestingly, as real-values. While a general feature map will yield complex pseudo-entropy values, we rigorously show that a certain class of feature maps that do not generate entanglement will always yield a positive real pseudo-entropy value. However, essentially for free, we get that a special unitary operator with symmetric eigenvalues will yield only positive real values.

The operator pseudo-entropy is mathematically derived, and we also rigorously derive its characteristics. A proposed metric to calculate the difference of two quantum operators is given. An application of three different data sets, two of which are synthetic and the other is real-world, to three well-applied encoding schemes is given. From further analysis of the experiments, we hypothesize that the concentrated distribution of unique values from the special unitary matrices from the singular value decomposition indicates an exponential concentration of values.      

The first concept of entanglement entropy was introduced by von Neumann, who extended the classical notion of entropy from probability measures to density matrices through the application of the trace operator \cite{nielsen2010quantum}. Considering the change of entanglement from one state to another, von Neumann entropy was extended to state-transition pseudo-entropy, where ``pseudo'' comes from the values of this entropy being complex \cite{nakata2020holographic}. Both concepts of entropy consider quantum states as the first step, which is an important value in physics. However, for quantum machine learning, it is the special unitary operator that generates the quantum state, and the operator is crafted then seeded by a real-valued data point. Thus, it is the gate operators that is the basis of the ``energy'' in the system, and is respective to individual data points. It is from this perspective that motivated the derivation of operator pseudo-entropy.

The rest of the paper is organized as follows. Section \ref{sec:pseudo-entropy} derives operator pseudo-entropy, a distance function to compare pseudo-entropy values, some characteristics of operator pseudo-entropy are rigorously derived. Section \ref{sec:analyze-encoding} moves into the comparison of operator pseudo-entropy, von Neumann entropy, and state-transition pseudo-entropy with three different data sets, two of which are synthetic, and three different quantum feature maps. Within this section we observe a direct connection between the exponential concentration of states, the entanglement of quantum states, and the distribution of unique values of the entries in the special unitary matrix from the spectral value decomposition of the operator. To display the robustness of pseudo-entropy, Section \ref{sec:comparison-other-methods} compares the technique against expressibility, expressivity, and symmetric encoding. Finally, Section \ref{sec:wrap-up} summarizes the results and proposes directions for future research.

\section{\label{sec:pseudo-entropy} Derivation of Operator Pseudo-Entropy}

For a basis for the definition of entropy of special unitary operators, we briefly described von Neumann entropy. Using the concept of classical entropy to quantum states, von Neumann defined the entropy of a density operator in the similar formula to that of Shannon entropy  \cite{vN18}. Specifically, for a density operator of the form $\displaystyle \mathcal{A} = \sum p_i \ket{i}\bra{i}$, where $p_i$ is the `frequency' weight for the eigen state $\ket{i}$, von Neumann entropy is defined as $S_{vN}( \mathcal{A} ) :=-\Tr\big[ \mathcal{A} \log(\mathcal{A}) \big] = -\sum_i p_i \log(p_i)$. In other words, the eigenvalues of the density matrix. 

For quantum circuits, it is the gates, or the special unitary operators, on the circuit that determine the energy in the system that generates the respective states and entanglement. Thus, these operators are our focus. Recall that an arbitrary special unitary operator $U$ has the singular value decomposition (SVD) of the form $U=V\Lambda V^{\dagger}$, where $V$ is special unitary and $\Lambda$ is a diagonal matrix where the entries are the eigenvalues of $U$. For completeness, the decomposition is a consequence of the Spectral Theorem \cite{hall2013quantum}, and this spectral decomposition is just a special case of singular value decomposition. Applying the operation of von Neumann entropy to a special unitary operator and using the properties of the trace, we see that 
\begin{equation}\label{eq:von-neumann-der}
\begin{split}
    & -\Tr\big( U\log(U) \big) = -\Tr\big( V\Lambda V^{\dagger} \log(V \Lambda V^{\dagger}) \big) \\
    & = -\Tr\big( V\Lambda V^{\dagger} V  \log(\Lambda ) V^{\dagger} \big) = -\Tr\big( V \Lambda \log(\Lambda ) V^{\dagger} \big) \\
    & = -\Tr\big( \Lambda \log(\Lambda ) V^{\dagger} V  \big) = -\Tr\big( \Lambda \log(\Lambda ) \big).
\end{split} 
\end{equation}
Since $U$ was arbitrary, we may define the pseudo-entropy of the operator as 
\begin{equation}\label{eq:pseudo-entropy}
    S(U) = -\frac{1}{\sqrt{\dim(U)}}\Tr\left(\Lambda \log(\Lambda) \right).
\end{equation}
Observe that the term $\dim(U)^{-1/2}$ normalizes the value since there will be linear growth of the values with respect to dimension of the special unitary matrix. For quantum algorithms, given the number of wires in a circuit $n$, then $\dim(U)^{-1/2} = 2^n$.

Notice that the definition of operator pseudo-entropy only changes the domain of arguments fed into the von Neumann entropy function. Furthermore, since the eigenvalues are the `energy' of the collection of quantum gates, this definition has a physical description, and are shown within this section. To see why this describes an 'energy', for the identity matrix $I$ it is clear that $S(I)=0$, which is intuitive since there are no transition of states.

For an initial analysis, for a special unitary $U$ recall that the eigenvalues are of the form $e^{\alpha i}$ for some $\alpha \in (- \pi, \pi)$. This yields
\begin{equation}\label{eq:basic-char}
\begin{split}
-\sqrt{\dim(U)}S(U) & = \sum_j e^{\alpha_j i} \log( e^{\alpha_j i} ) \\
& = \sum_j \big( \cos(\alpha_j) + i\sin(\alpha_j) \big)\alpha_j i \\
& = i\sum_j \alpha_j  \cos(\alpha_j) - \sum_j \alpha_j  \sin(\alpha_j).
\end{split} 
\end{equation}
Consequently, $-\Tr\big( \Lambda \log(\Lambda ) \big) \in \mathbb{R}$ if and only if $\displaystyle \sum_j \alpha_j \cdot \cos(\alpha_j) = 0$, that is, when $S(\cdot)$ is well-defined. Operator pseudo-entropy being well-defined is discussed below.   

Given that the complex unit circle is the domain of this entropy function, one needs to incorporate the branch-cut, ensuring this concept of pseudo-entropy is well-defined. We choose the standard branch cut for the complex logarithm, which is the negative real axis. Hence, the principal branch is $(-\pi,\pi)$ \cite{NTTTW21}. This branch is exactly the branch of the logarithm in the matrix method in SciPy \cite{2020SciPy-NMeth}, which is essential for experiments in Section \ref{subsec:experiments}. With this principal branch, the definition is well-defined and continuous for all eigenvalues that are not $-1$. One may consider $-1$ in the domain, however, this needs to be carefully considered but would make the pseudo-entropy function not continuous at this point. We propose to use $e^{-\pi}$ as the value at $-1$.

\begin{figure*}[th!]
     \centering
     \begin{subfigure}{.32\textwidth}
    \centering
    \includegraphics[width=.9\textwidth]{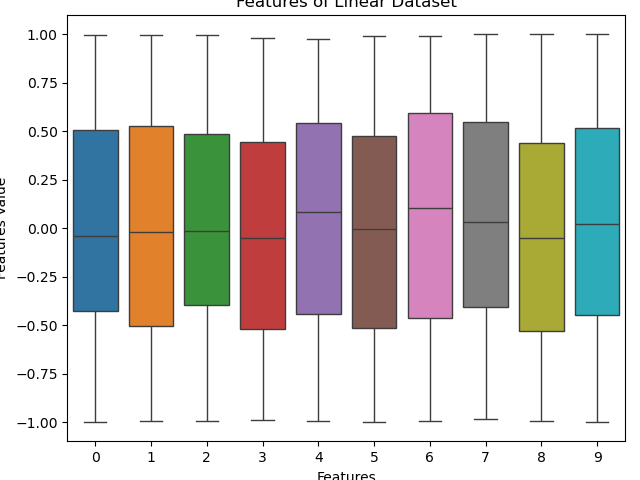}\caption{Linearly separable synthetic data.}\label{subfig:linear-sep}
     \end{subfigure}
     \hfill
     \begin{subfigure}{.32\textwidth}
    \centering
    \includegraphics[width=.9\textwidth]{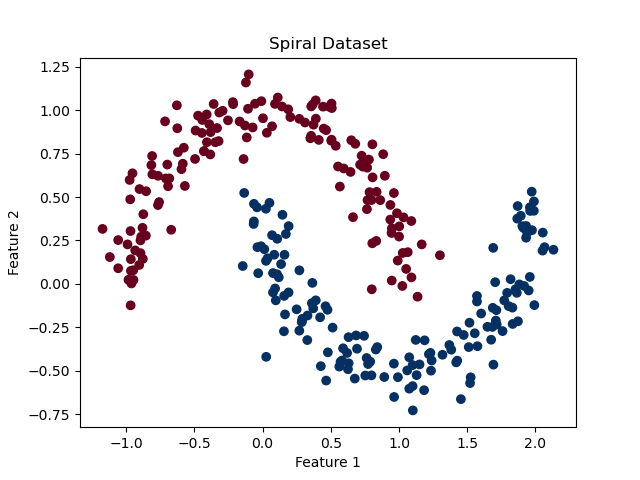}\caption{Synthetic spiral data.} \label{subfig:spiral-data}
     \end{subfigure}
     \hfill
     \begin{subfigure}{.32\textwidth}
    \centering
    \includegraphics[width=.9\textwidth]{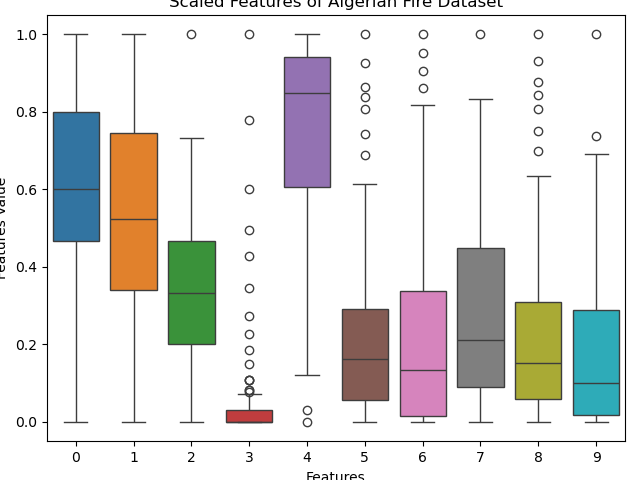}\caption{Algerian fire scaled data.} \label{subfig:trans-state-data-example}
     \end{subfigure}
\caption{Visualization of the features for each data set.}
\label{fig:trans-state-pseudo-entropy-data}
\end{figure*}

We now show that if the quantum circuit does not generate entanglement then the pseudo-entropy is real and positive. However, it is not true that if pseudo-entropy is real then there is no entanglement. To see why this would be true, as shown from Equation \ref{eq:basic-char}, for each complex eigenvalue $\lambda_i$ assume there exists an eigenvalue $\lambda_j$ with $i \neq j$, such that $\lambda_i = \overline{\lambda_j}$, then the pseudo-entropy value is real and positive. We call an operator with this characteristic \define{symmetric}. Symmetry is quite natural for a circuit with no entanglement, and this is shown in the proposition below. However, as displayed in Section \ref{subsec:experiments}, there are instances when entanglement is present but all values are real. 

For the following proposition, notice that we assume special unitary operators and not unitary gates, in general, like Hadamard and phase. Recall that unitary operators are special unitary operators but with a global phase. As this phase disappears in the measurement, such gates are considered as a good approximation.

\begin{proposition}\label{prop:pseudo-entropy-real-values}
    If $U$ represents a special unitary operator acting only on single qubits in a circuit and $-1$ is not an eigenvalue, then $S(U) \in \mathbb{R}_{\geq 0}$. Furthermore, for general special unitary operators $U_1$ and $U_2$, if $-1$ is not an eigenvalue for either operator and the addition of exponents  from an arbitrary eigenvalue of $U_1$ and an arbitrary eigenvalue of $U_2$ do not cross over the principal branch then
\begin{equation}\label{eq:tensor-entropy-equality}
  S\left( U_1 \bigotimes U_2 \right) = \Tr(U_2) \cdot S(U_1) + \Tr(U_1) \cdot S(U_2). 
\end{equation}  
\end{proposition}
\begin{proof}
For a circuit with a single wire, ergo $U$ is a $2 \times 2$ special unitary operator, the eigenvalues of $U$ are complex conjugates, see  Chapter 1 in Hall \cite{hall2015lie}. Thus $S(U) \in \mathbb{R}_{\geq 0}$. For $\displaystyle U = \bigotimes_i U_i$ the special unitary operator of a circuit with an arbitrary finite number of wires, note that, for the SVD $U_i = V_i \Lambda_i V_i^{\dagger}$,
\begin{equation} \label{eq:kron-decomp}
    S\left( \bigotimes_i U_i \right) = S\left( \bigotimes_i \Lambda_i \right).
\end{equation}
We claim that given an eigenvalue of $U$, the complex conjugate of this eigenvalue is also an eigenvalue of $U$. By Equation \ref{eq:kron-decomp}, we will focus on the diagonal matrix of eigenvalues. Observe that the eigenvalues of $U_1\bigotimes U_2$ are 
\begin{equation*}
    \begin{split}
        \Lambda_1 \bigotimes \Lambda_2 & = \begin{bmatrix} e^{i \theta_1} & 0 \\
        0& e^{ -i \theta_1} \end{bmatrix} \bigotimes \begin{bmatrix} e^{i \theta_2} & 0 \\
        0& e^{ -i \theta_2} \end{bmatrix} \\
        & = \begin{bmatrix}  e^{i \theta_1} e^{i \theta_2} & 0 & 0 & 0\\
        0 & e^{i \theta_1} e^{ -i \theta_2} & 0 & 0 \\
        0 & 0 & e^{ -i \theta_1} e^{i \theta_2} & 0\\
        0 & 0 & 0 & e^{ -i \theta_1} e^{ -i \theta_2} \\ \end{bmatrix},
    \end{split}
\end{equation*}
which clearly holds the claim. Induction finishes the claim. Therefore, $S(U) \in \mathbb{R}_{\geq 0}$ for a general circuit with single qubit operators. For multiple single qubit operators on a circuit, say $A_1,\ldots,A_n\in \mathrm{SU}(2)$, there exists a $C \in \mathrm{SU}(2)$ such that $A_1 \cdot\ldots\cdot A_n = C$. Therefore, the claim holds.  

From Equation \ref{eq:kron-decomp}, with the decomposition $U_1 = V_1 \Lambda_1 V_1^{\dagger}$ and $U_2 = V_2 \Lambda_2 V_2^{\dagger}$, we have the equality $\Tr\Big( U_1 \bigotimes U_2 \log(U_1 \bigotimes U_2) \Big) = \Tr\Big( \Lambda_1 \bigotimes \Lambda_2 \log(\Lambda_1 \bigotimes \Lambda_2) \Big)$. Taking into account the logarithm, we see that $\log(\Lambda_1 \bigotimes \Lambda_2) = \log(\Lambda_1 \bigotimes I \times I \bigotimes \Lambda_2) = \log(\Lambda_1 \bigotimes I) + \log(I \bigotimes \Lambda_2)$, since the matrices commute. Now, since the matrices are diagonal, $\log(\Lambda_1 \bigotimes I) = \log(\Lambda_1) \bigotimes I$ and $\log(I \bigotimes \Lambda_2) = I \bigotimes \log(\Lambda_2)$. The rest of the claim now follows. 
\end{proof}

Given that general operator pseudo-entropy values are complex, Proposition \ref{prop:pseudo-entropy-real-values} precisely shows that when entanglement is not introduced into a circuit there will never be a complex value. Moreover, observe that Equation \ref{eq:tensor-entropy-equality} is a generalized version of a characteristic of von Neumann entropy. 

\begin{remark}
    From the concept of a symmetric operator and the proof of the proposition \ref{prop:pseudo-entropy-real-values}, we observe that operators with asymmetric eigenvalues transform the data in a highly nonlinear manner. Therefore, the imaginary value of pseudo-entropy displays the intensity of this nonlinearity.
\end{remark}

\begin{figure*}[!t]
    \centering
    \renewcommand{\arraystretch}{1.}%
    \begin{tabularx}{\linewidth}{cccc} 
     & Linearly Separable &  Spiral & Algerian Fire \\
     \adjustbox{raise = 6 ex}{ \rotatebox{90}{\stackanchor[3pt]{Pseudo-Entropy}{Real Values}} }
    & \includegraphics[height = 3.2cm]{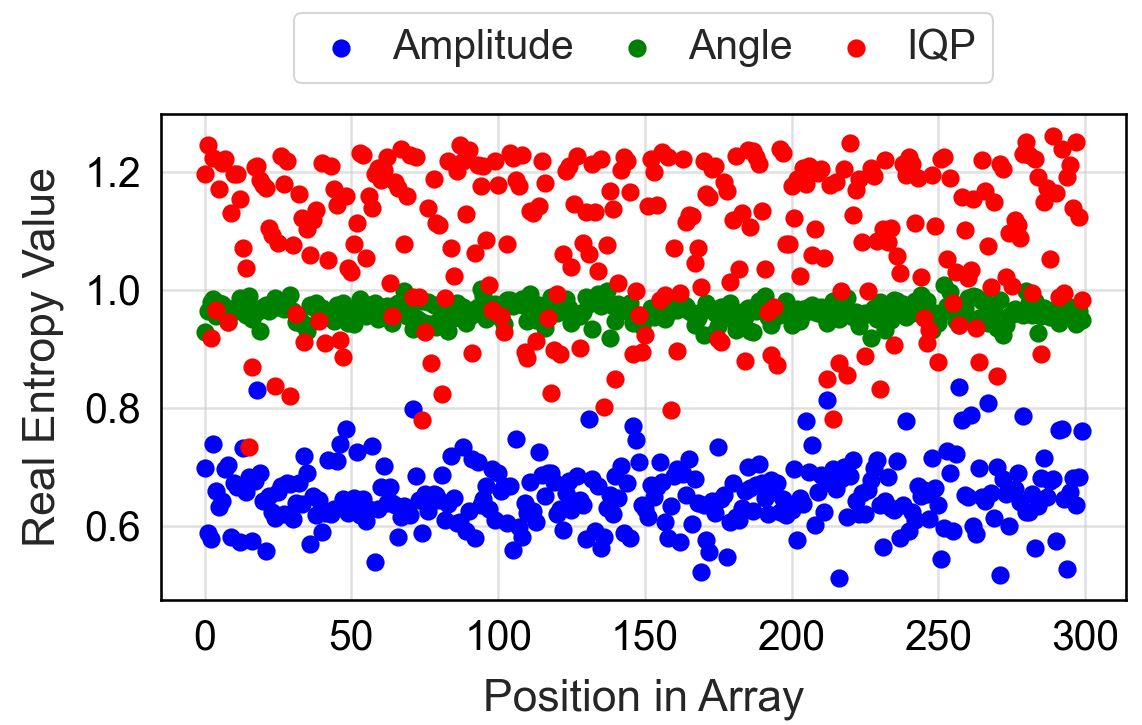} 
    & \includegraphics[height = 3.2cm]{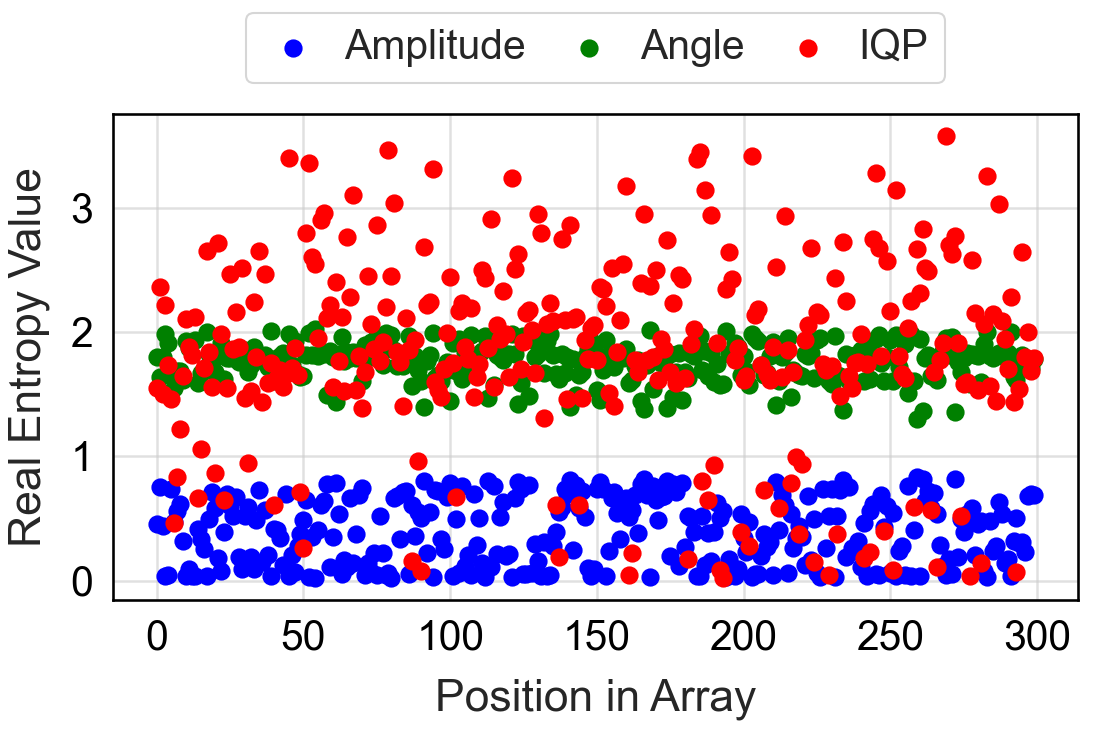}
    & \includegraphics[height = 3.2cm]{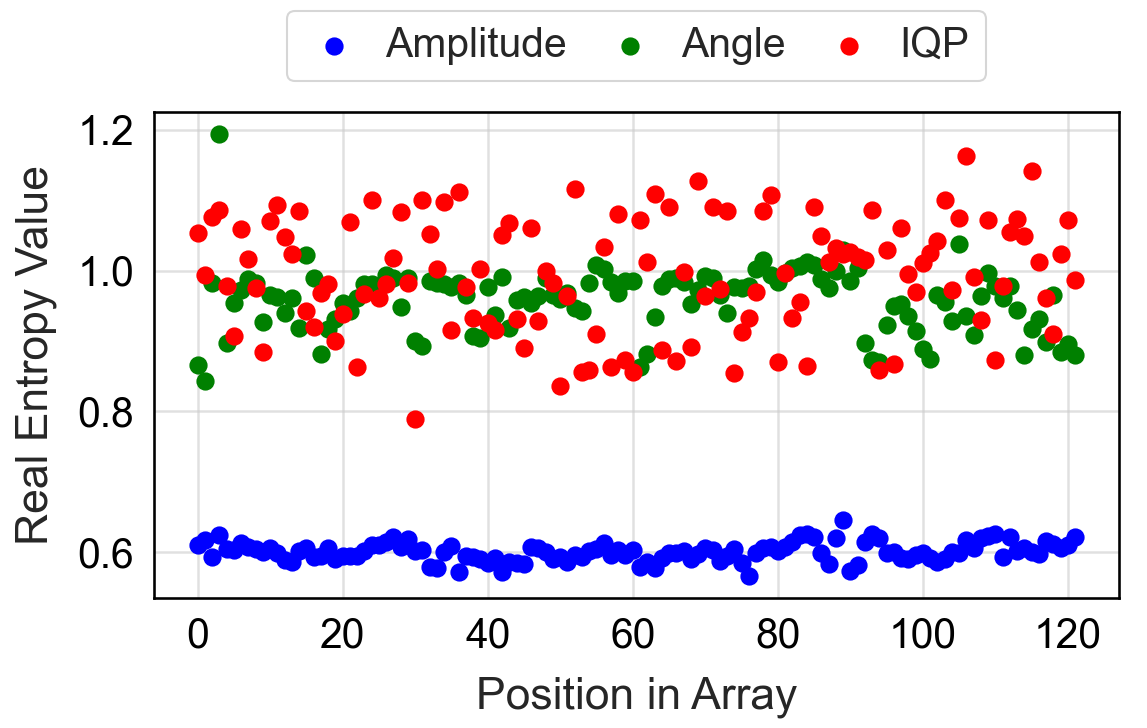}
    \\ \adjustbox{raise = 5 ex}{ \rotatebox{90}{\stackanchor[3pt]{Pseudo-Entropy}{Imaginary Values}} }
    &   \includegraphics[height = 3.5cm]{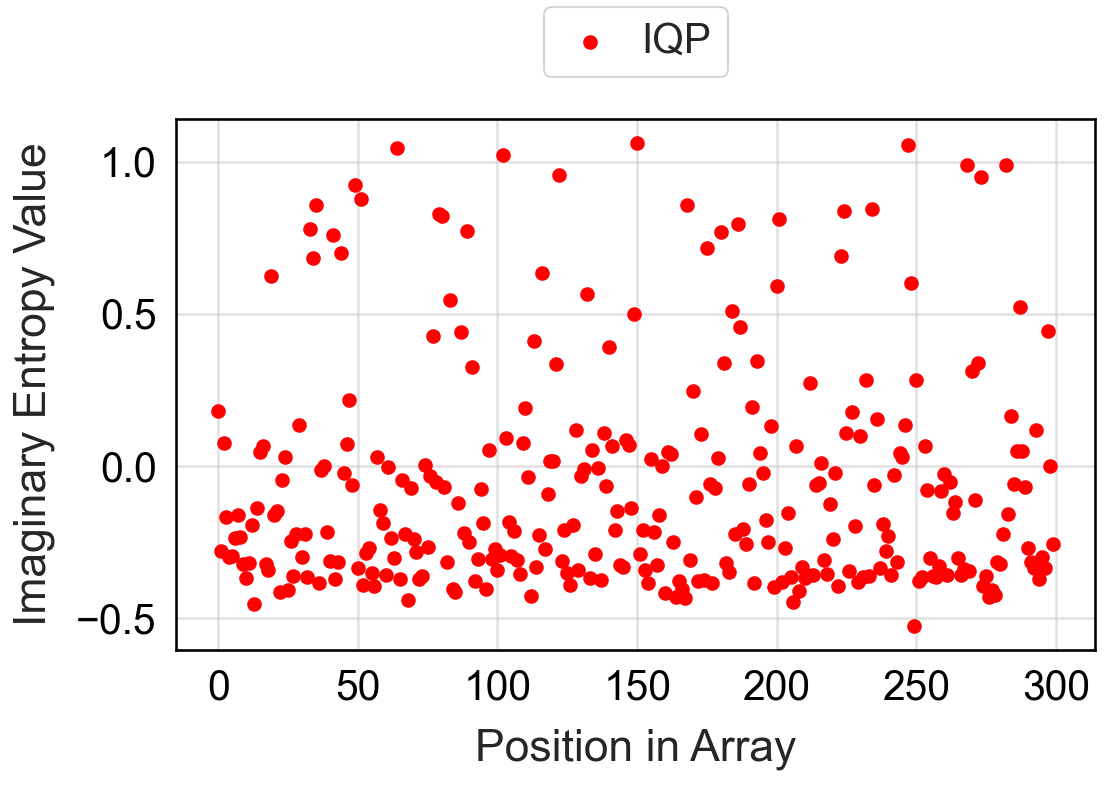}
    &  \includegraphics[height = 3.5cm]{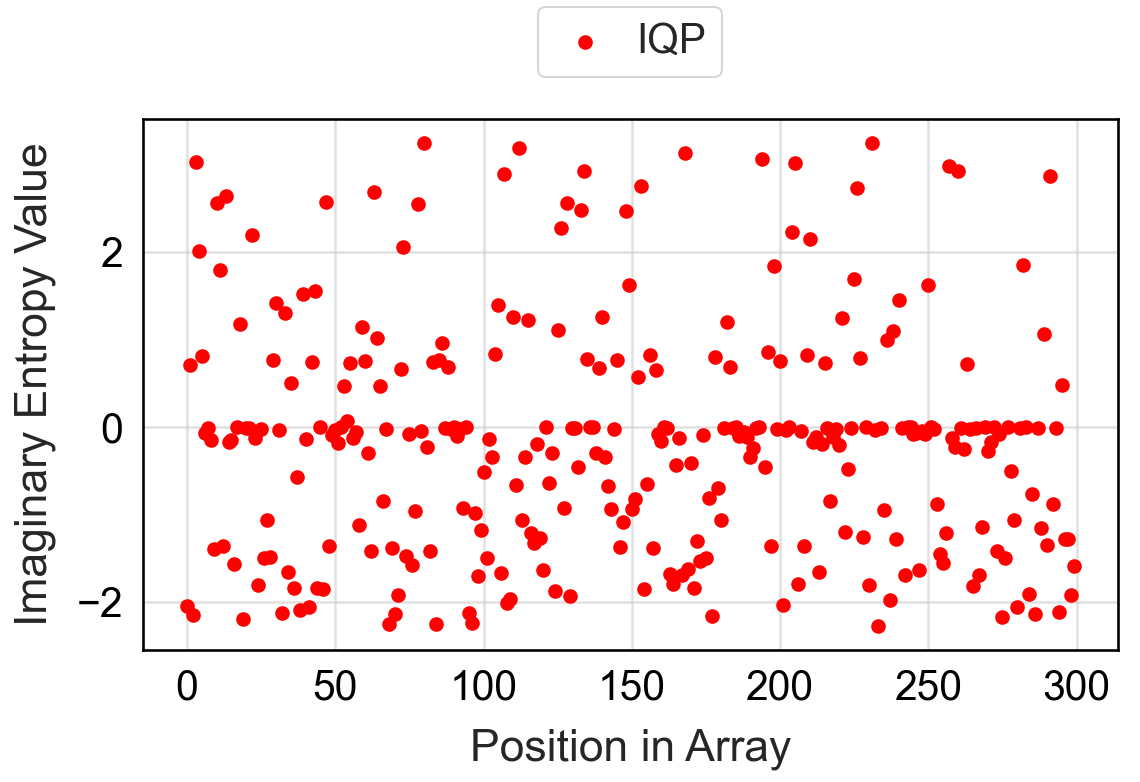}
    & \includegraphics[height = 3.5cm]{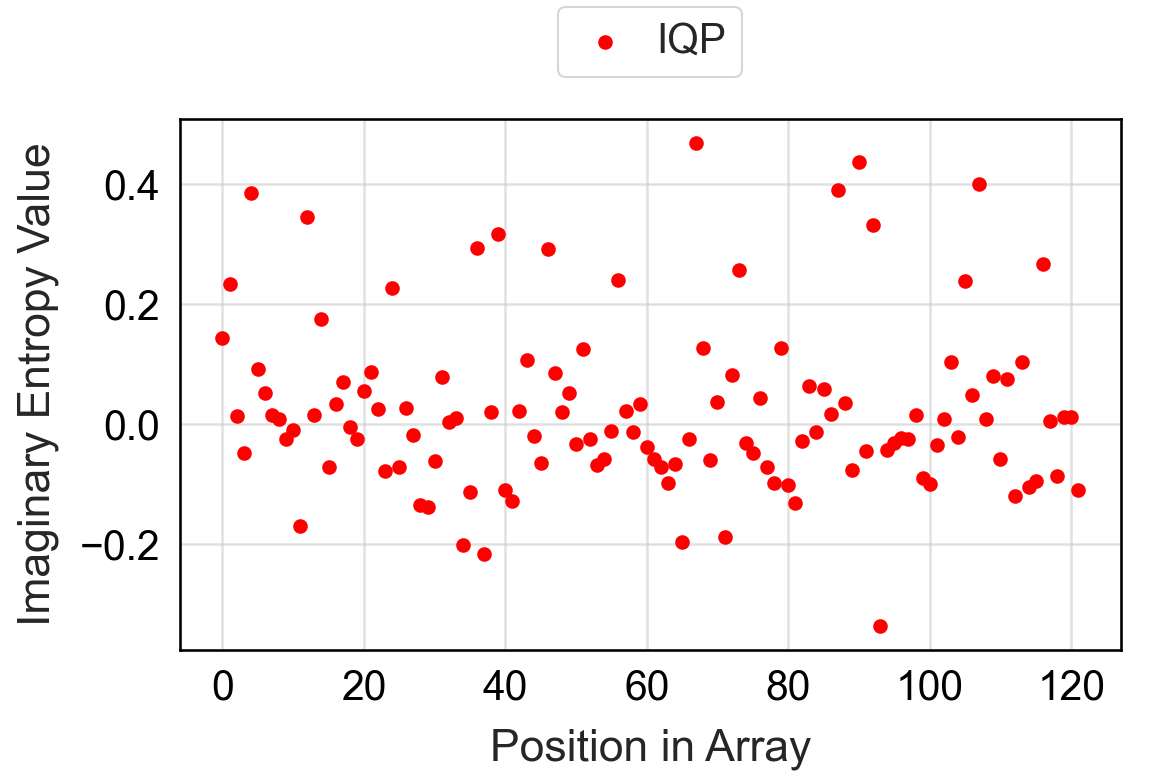} 
    \\ 
    \adjustbox{raise = 6.5 ex}{ \rotatebox{90}{\stackanchor[3pt]{von Neumann}{Entropy}} } 
    &   \includegraphics[height = 3.2cm]{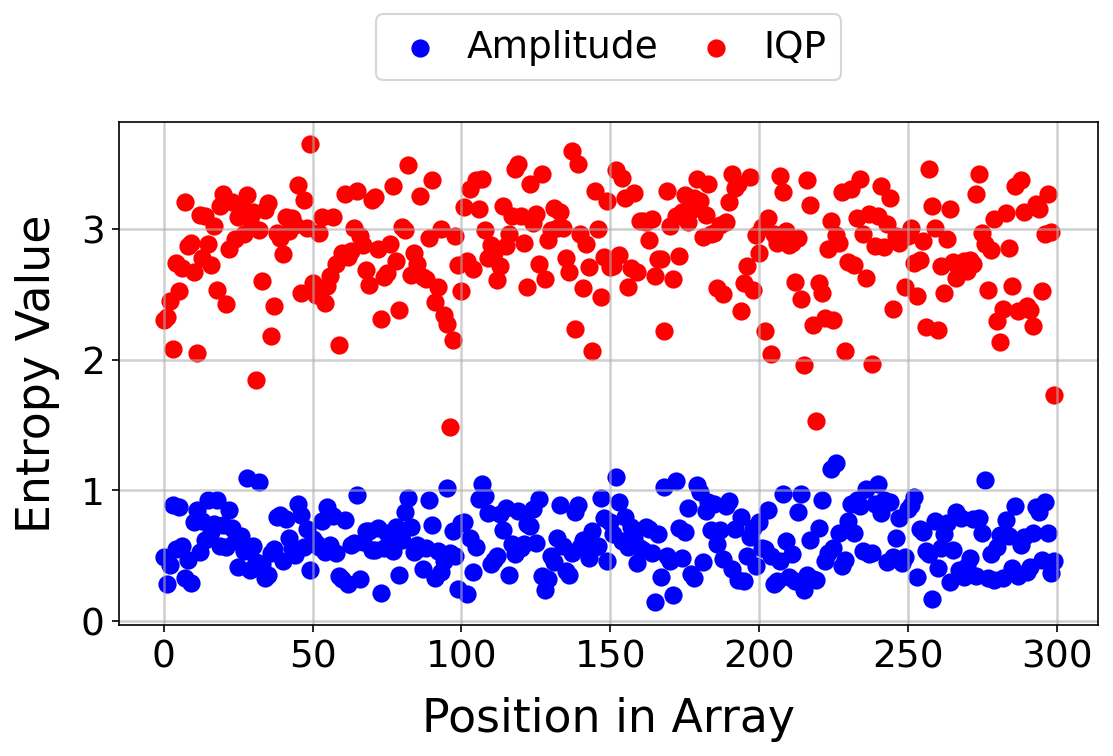}
    &  \includegraphics[height = 3.2cm]{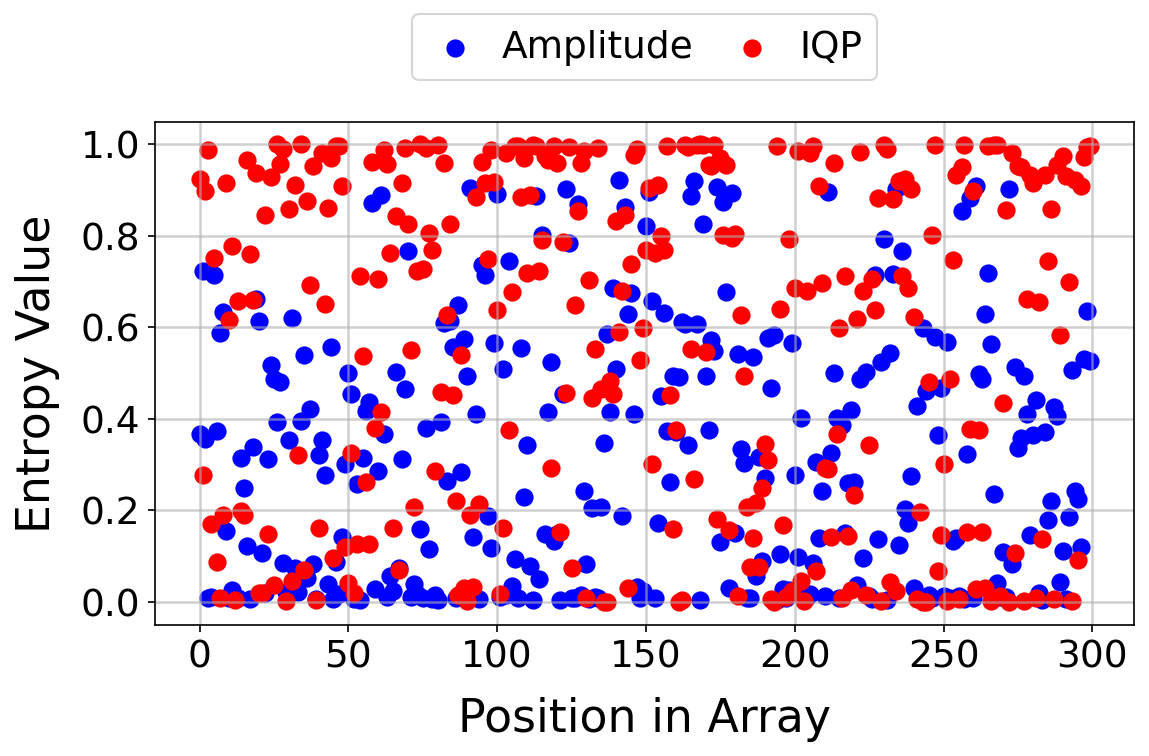}
    & \includegraphics[height = 3.2cm]{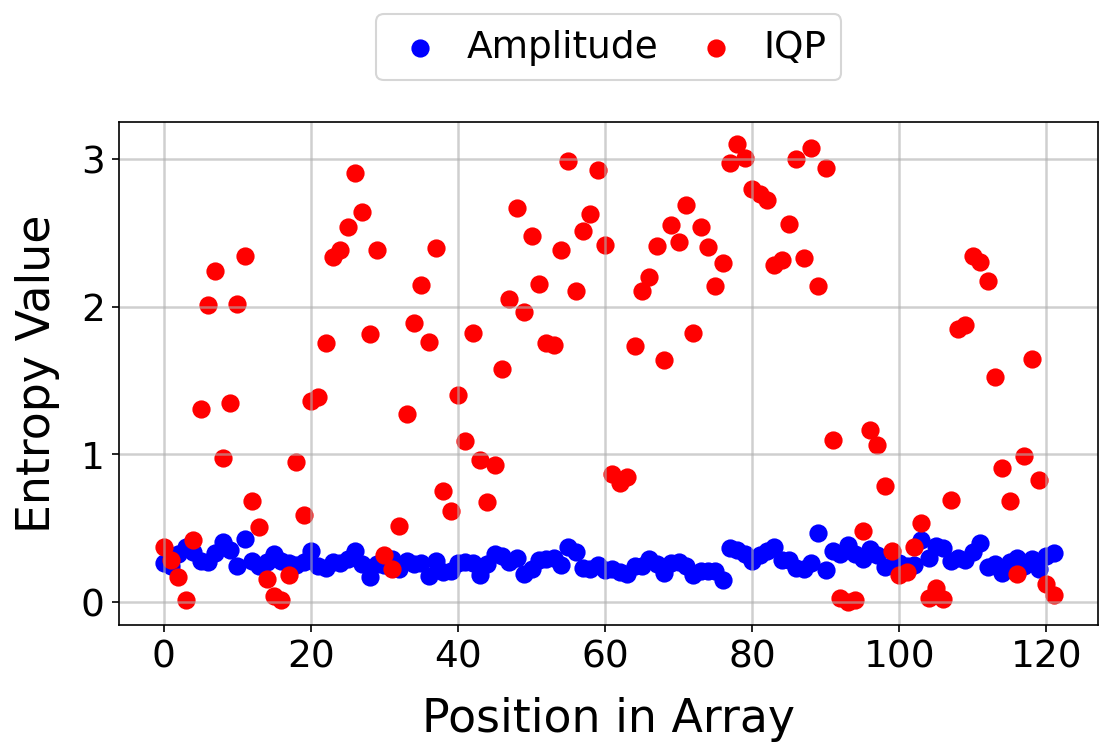} 
    \\ 
     \adjustbox{raise = 4.5 ex}{ \rotatebox{90}{\stackanchor[3pt]{State-Transition}{Real Values}} }
    & \includegraphics[width = 4.7cm]{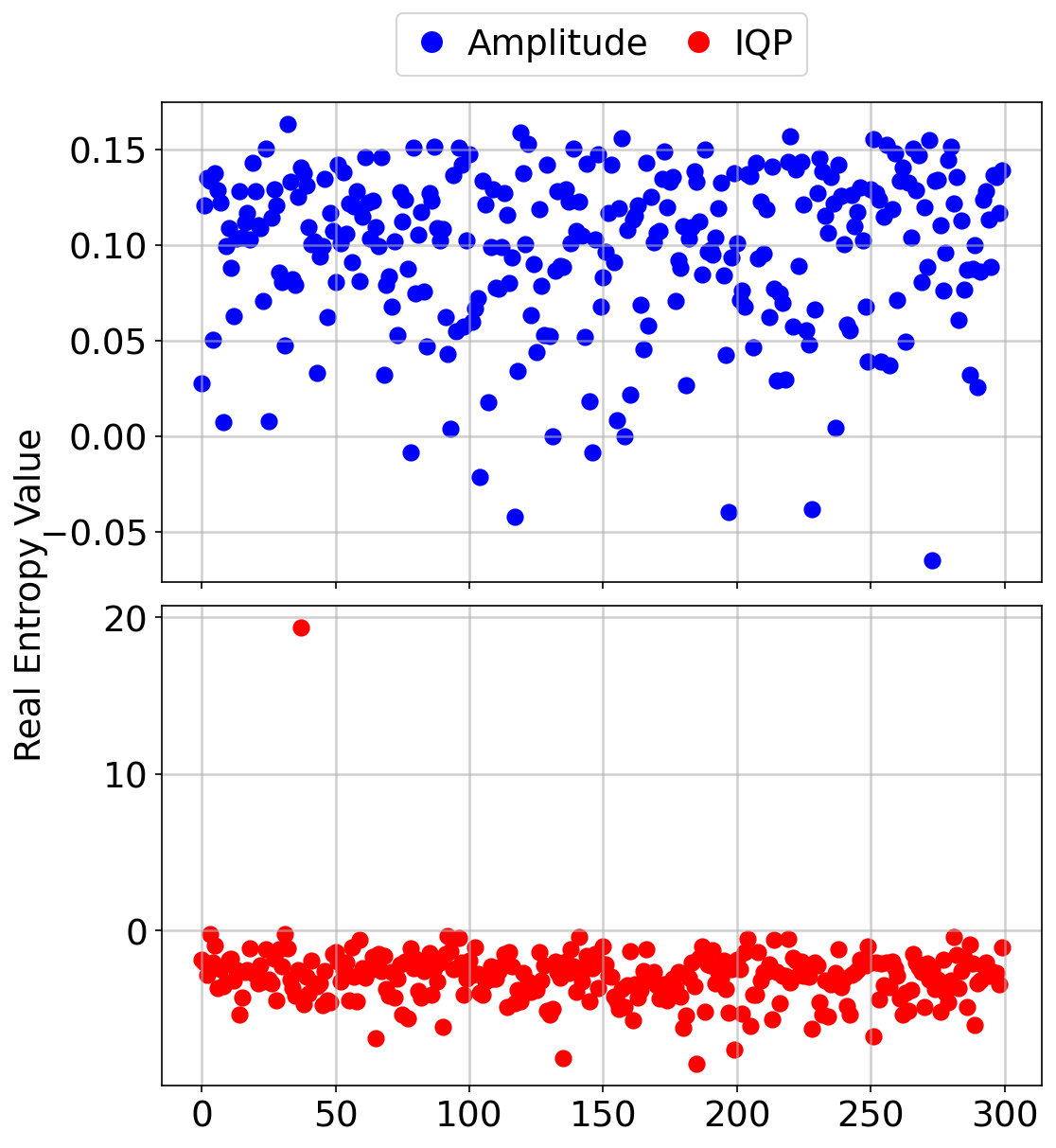} 
    & \includegraphics[width = 4.7cm]{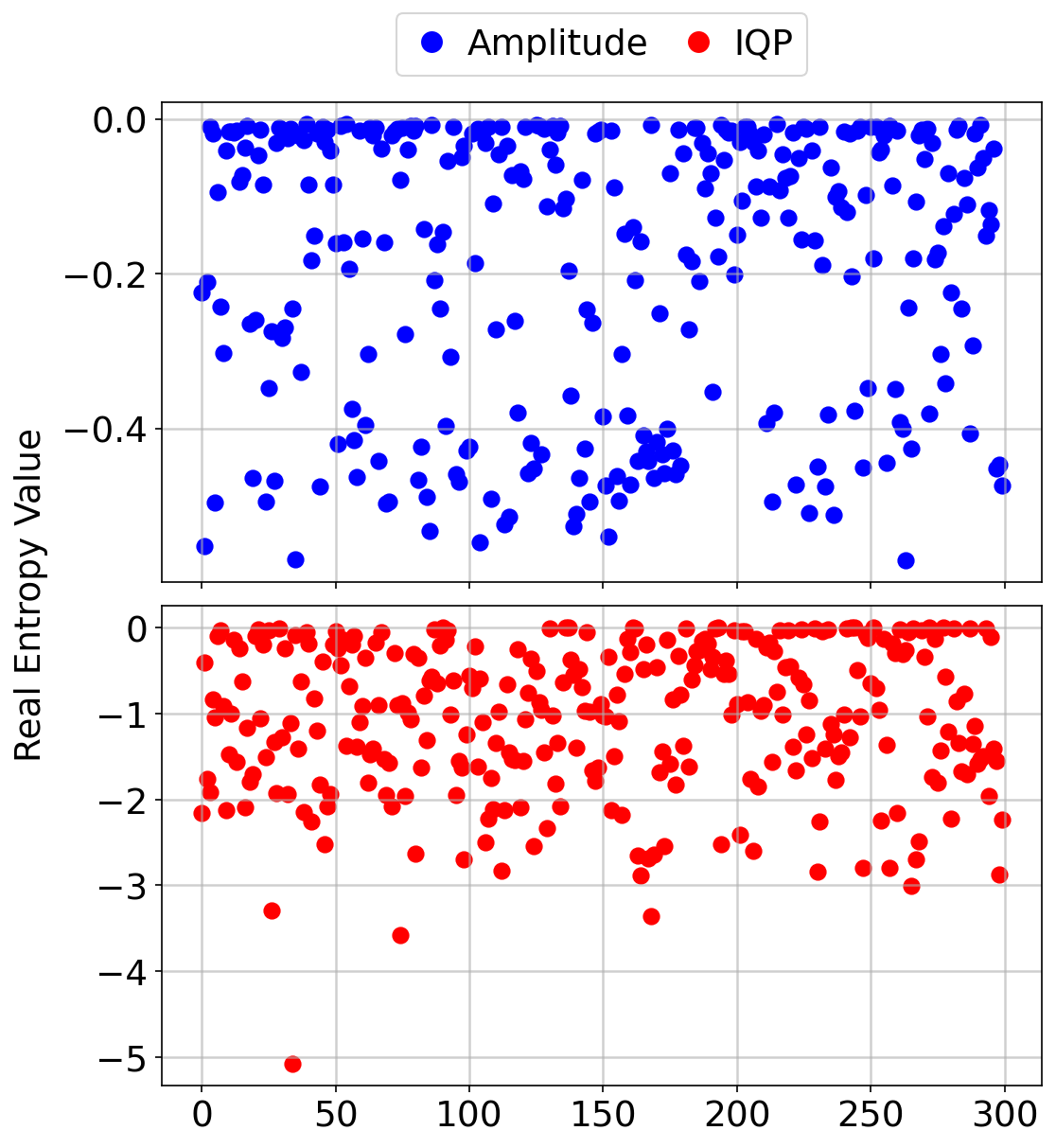}
    & \includegraphics[width = 4.7cm]{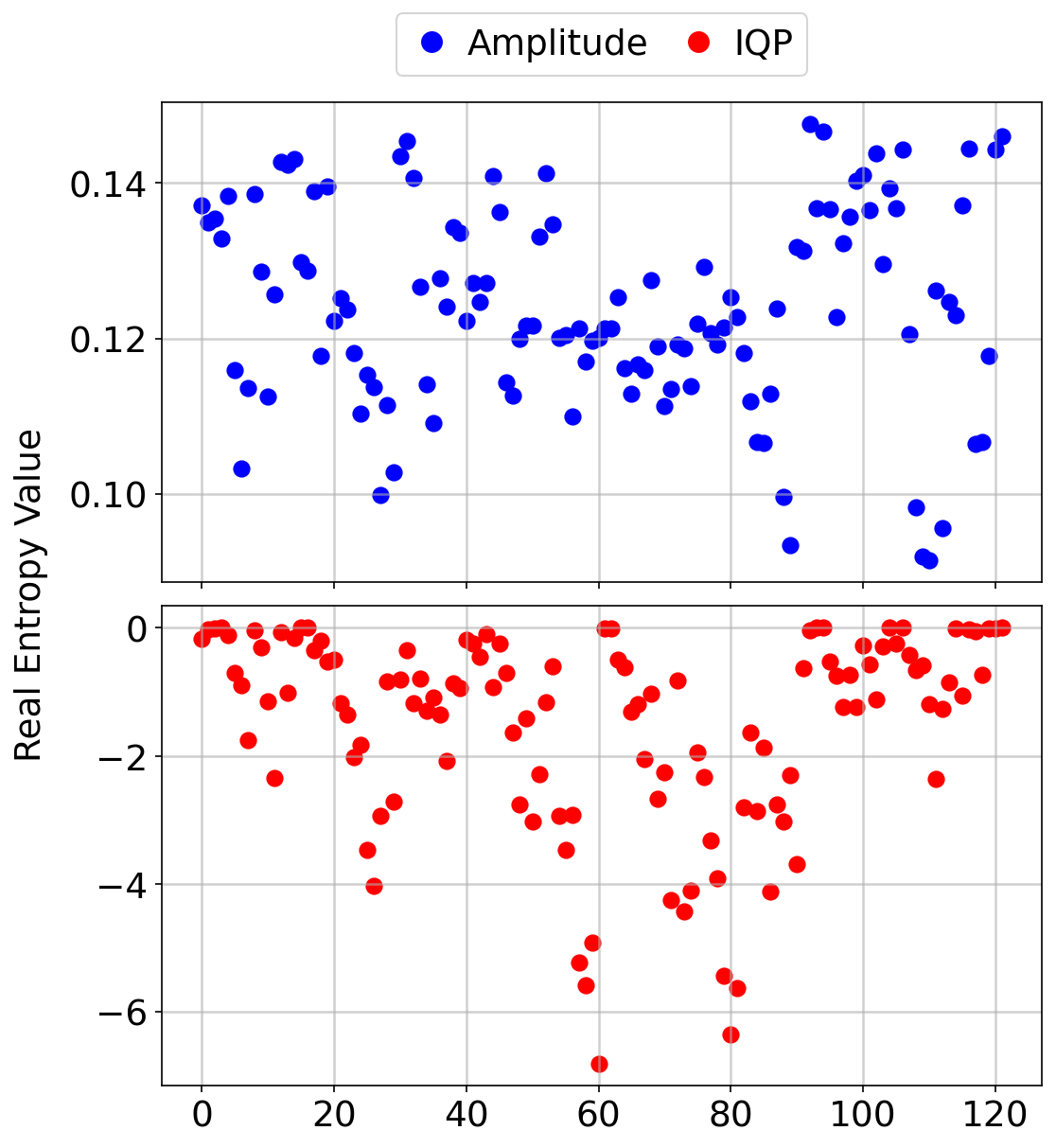}
    \\ 
    \adjustbox{raise = 4 ex}{ \rotatebox{90}{\stackanchor[3pt]{State-Transition}{Imaginary Values}} } 
    &   \includegraphics[width = 4.7cm]{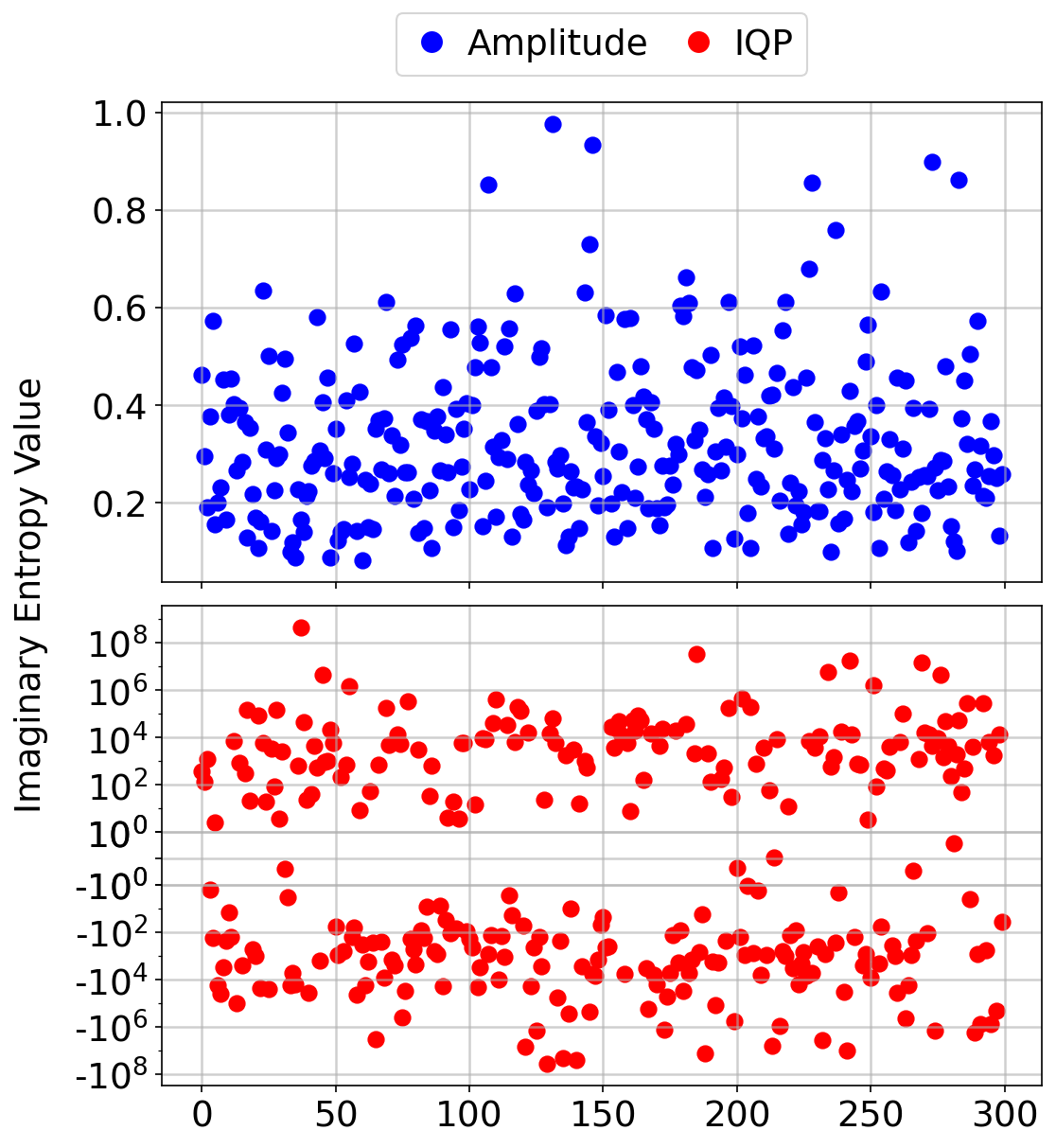}
    &  \includegraphics[width = 4.7cm]{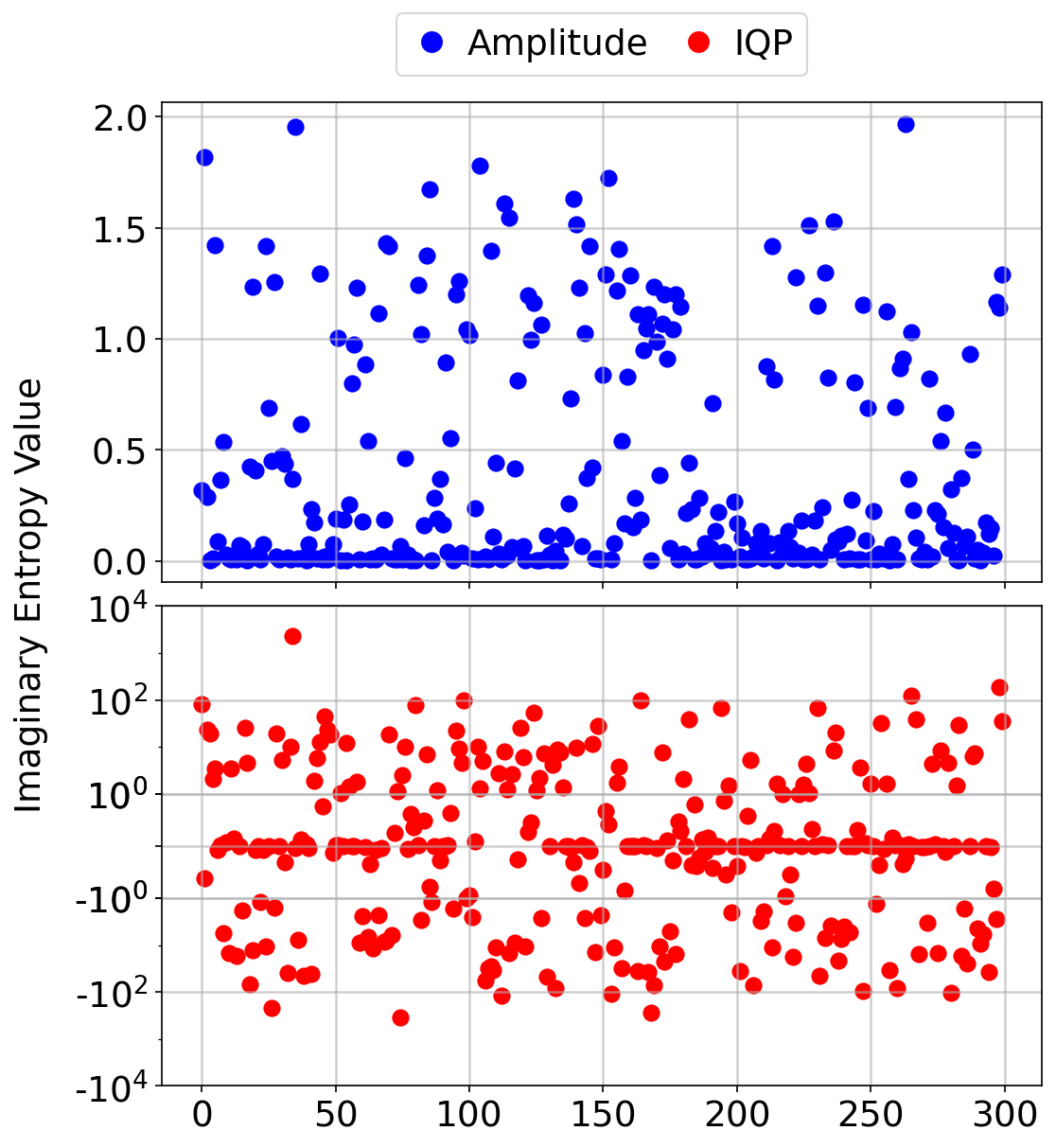}
    & \includegraphics[width = 4.7cm]{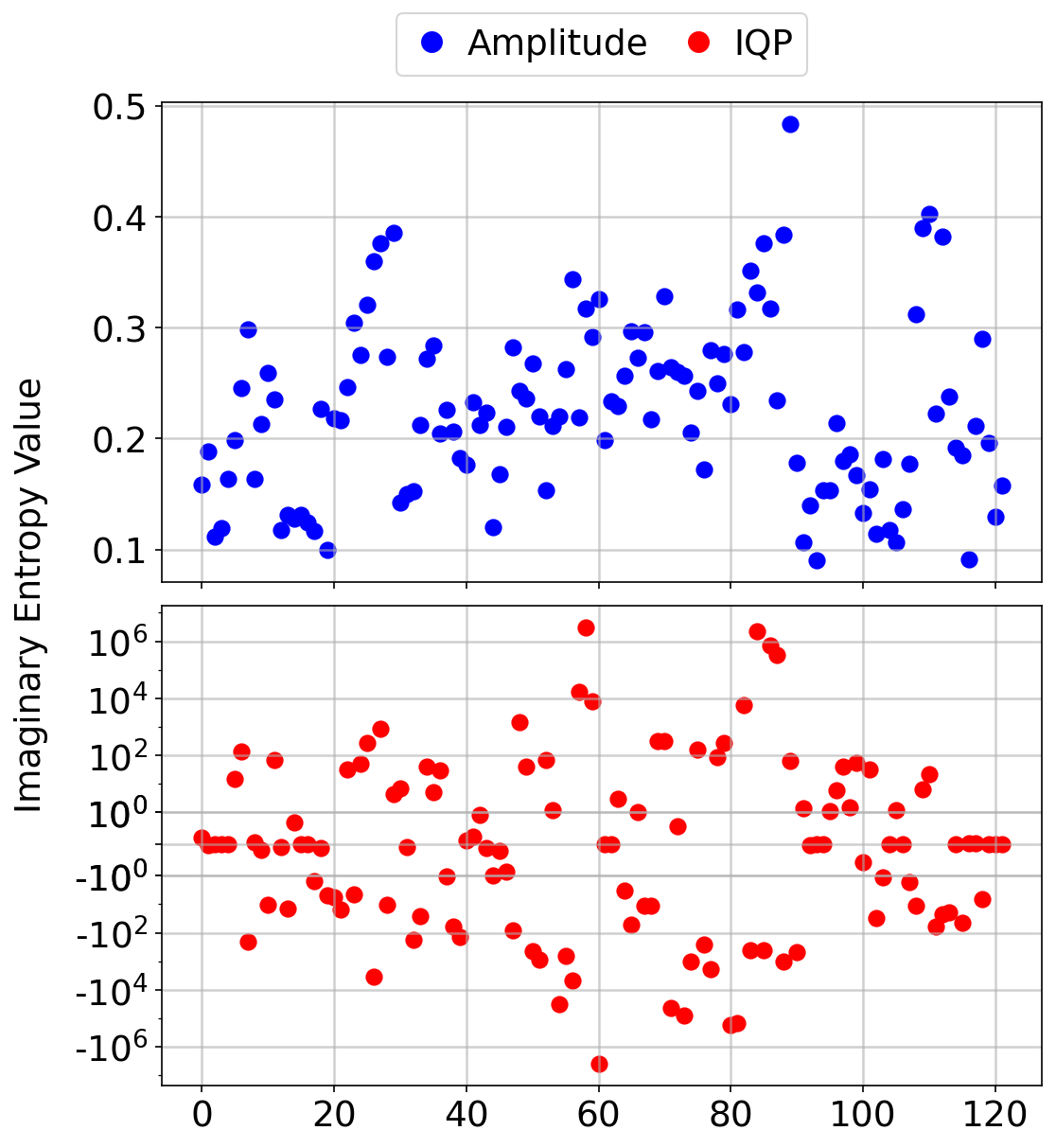} 
    \end{tabularx}
    \caption{Operator pseudo-entropy, von Neumann entropy, and the average state-transition pseudo-entropy values for three different data sets with the Angle encoding, Amplitude encoding, and IQP encoding schemes. The IQP averaged state-transition pseudo-entropy values are quite large, and to adjust the imaginary values have been log-scaled and the scatter plots are separated for a more granular visual analysis. The entropy values for the Angle encoding scheme are excluded since all values are zero. }
    \label{fig:pseudo-entropy-example-three-dataset}
\end{figure*}

\subsection{Distance Function of Operator Pseudo-Entropy}
With the definition of operator pseudo-entropy, it is not obvious how one would calculate the mutual information between two operators. Moreover, it would interesting to calculate the mutual information from the growth of a circuit, either with additional gates or additional qubits. We answer this question by defining the distance between the pseudo-entropy of operators, but first we motivate how this definition was derived.   

Considering entanglement as nodes from one wire, or state, to another and taking that respective graph, Qi \cite{qi6282exact} defines the length of between sites $i$ and $j$ as $\displaystyle -\log\Big( \frac{I(i:j)}{2\log(D)} \Big)$, where $I(i:j)$ is von Neumann mutual information, and $D$ is the number of states of $i$ and the term $2\log(D)$ is shown to be an upper bound. For completeness, Hyatt, Garrison, and Bauer \cite{hyatt2017extracting} posit that $2\log(D) = 2\log(2)$, and Anand et al. \cite{anand2024information} further extend this definition and posit that the normalization term $2\log(D)$ should be $4\log(2)$. 

To compare the pseudo-entropy of two special unitary operators we apply the concept in Qi \cite{qi6282exact}. Take $U$ acting on the Hilbert spaces, adjusting for small pseudo-entropy values near zero and large pseudo-entropy values, we define $\widehat{ S }(U) := \displaystyle \frac{ S(U) + 1 }{ \dim(U)^{1/2}\log(2) }$. Thus, the `distance' would then be $\displaystyle d_{U} := - \log\left( \widehat{ S }(U) \right)$.

\begin{definition}\label{def:diff-metric}
We define the difference of the pseudo-entropy of two special unitary operators, $U_1$ and $U_2$, as 
\begin{equation*}
    \begin{split}
        & \Dist(U_1,U_2) = d_{U_1} - d_{U_2}  =  \log\left( \frac{\widehat{ S }(U_2)}{\widehat{ S }(U_1)} \right) \\
        & = \log\left( S(U_2) + 1 \right) - \log\left( S(U_1) + 1 \right) \\ 
        &  + \frac{\log\big(\dim(U_1)\big) - \log\big(\dim(U_2)\big)}{2}. 
    \end{split}
\end{equation*}
Observe that $\Dist(U_1,U_2) = - \Dist(U_2,U_1)$ and when the modulus is applied to the function this definition holds the properties of a mathematical metric.
\end{definition}

\subsection{State-Transition Pseudo-Entropy}

The concept of generalizing von Neumann entropy to operators beyond density matrices is fairly recent \cite{vedral1998entanglement,kitaev2006topological,hu2019generalized,nakata2020holographic,mukherjee2022pseudo}. A particularly relevant construction is the \emph{state-transition pseudo-entropy} \cite{nakata2020holographic}, which is defined in terms of the transition operator between two pure quantum states. For two non-orthogonal states $\ket{\varphi}$ and $\ket{\psi}$, the transition operator is
\begin{equation}\label{eq:transition-operator}
    \mathcal{T}^{\varphi|\psi} = \frac{ \ket{\varphi}\bra{\psi} }{ \braket{\varphi}{\psi} }.
\end{equation}
In general, $\mathcal{T}^{\varphi|\psi}$ is not Hermitian and does not represent a physical density matrix. Nevertheless, one may apply the von Neumann entropy functional to this operator to obtain the state-transition pseudo-entropy
\begin{equation}\label{eq:stpe}
    S_{ST}\!\left( \mathcal{T}^{\varphi|\psi} \right) := - \Tr\!\left( \mathcal{T}^{\varphi|\psi} \log \mathcal{T}^{\varphi|\psi} \right).
\end{equation}
In general, however, $S_{ST}$ can take complex values, with its real part and magnitude reflecting the degree of entanglement encoded in $\ket{\varphi}$ relative to $\ket{\psi}$.

To study entanglement across subsystems, consider a bipartition of the Hilbert space $\mathcal{H} = \mathcal{H}_A \otimes \mathcal{H}_B$. One defines the reduced transition operator by tracing out subsystem $B$,
\begin{equation}\label{eq:reduced-transition}
    \mathcal{T}_A^{\varphi|\psi} = \Tr_B\!\left( \frac{\ket{\varphi}\bra{\psi}}{\braket{\varphi}{\psi}} \right),
\end{equation}
and correspondingly the reduced state-transition pseudo-entropy
\begin{equation}\label{eq:reduced-stpe}
    S_{ST}(A) = - \Tr\!\left( \mathcal{T}_A^{\varphi|\psi} \log \mathcal{T}_A^{\varphi|\psi} \right).
\end{equation}
This reduced quantity generalizes the usual bipartite entanglement entropy by applying the von Neumann entropy functional to transition operators rather than density matrices.

In the setting of a layered quantum circuit $U = U_L \cdots U_1$, one can follow the dynamics of pseudo-entropy across successive layers. Denoting the intermediate states
\begin{equation}
    \ket{\psi^{(\ell)}} = U_\ell \cdots U_1 \ket{0}^{\otimes n}, \quad \ell = 0,1,\dots,L,
\end{equation}
with $\ket{\psi^{(0)}} = \ket{0}^{\otimes n}$, the reduced transition operator between consecutive layers is
\begin{equation}\label{eq:layer-transition}
    \mathcal{T}_A^{(\ell|\ell-1)} = \Tr_B\!\left( \frac{\ket{\psi^{(\ell)}}\bra{\psi^{(\ell-1)}}}{\braket{\psi^{(\ell-1)}}{\psi^{(\ell)}}} \right).
\end{equation}
The corresponding layer pseudo-entropy is
\begin{equation}\label{eq:layer-stpe}
    S_{ST}^{(\ell)}(A) = - \Tr\!\left( \mathcal{T}_A^{(\ell|\ell-1)} \log \mathcal{T}_A^{(\ell|\ell-1)} \right).
\end{equation}
Finally, to capture the typical entanglement structure across the entire circuit, one defines the averaged pseudo-entropy
\begin{equation}\label{eq:avg-stpe}
    \overline{S}_{ST}(A) = \frac{1}{L} \sum_{\ell=1}^L S_{ST}^{(\ell)}(A).
\end{equation}
This averaged quantity provides a compact metric for comparing entanglement growth and distribution across different circuits or input states.

\subsection{Qualitative Comparison Between Entropy Methods}
To calculate quantum entropy, in general, all techniques are just the same functional applied to a different domain of matrices. However, the information extracted are all quite different. The entropy of von Neumann measures the randomness or mixedness of a given quantum state as described by a density matrix, which aggregates information. This value is always real and nonnegative. State-transition pseudo-entropy describes the change in entanglement between states, giving more granular information than von Neumann. The general value is complex and entanglement has to be present within the quantum state, otherwise, the value is zero. Operator pseudo-entropy only considers the operator and not the quantum states, and describes both superposition and entanglement. Moreover, non-symmetric entanglement is given by the imaginary value of the entropy. Yielding more granular information than von Neumann, but not as granular as state-transition. However, state-transition pseudo-entropy requires considerably more computational resources and considerably more time to compute to calculate over operator pseudo-entropy.  

The following section goes deeper into the comparison through numerical analysis from three different data sets, coupled with three different encoding schemes. 

\section{\label{sec:analyze-encoding} Analyzing Feature Maps}

\subsection{\label{subsec:analysis} The Difficulty of Analyzing Feature Maps}

\begin{figure}[th!]
     \centering
    \includegraphics[width=.35\textwidth]{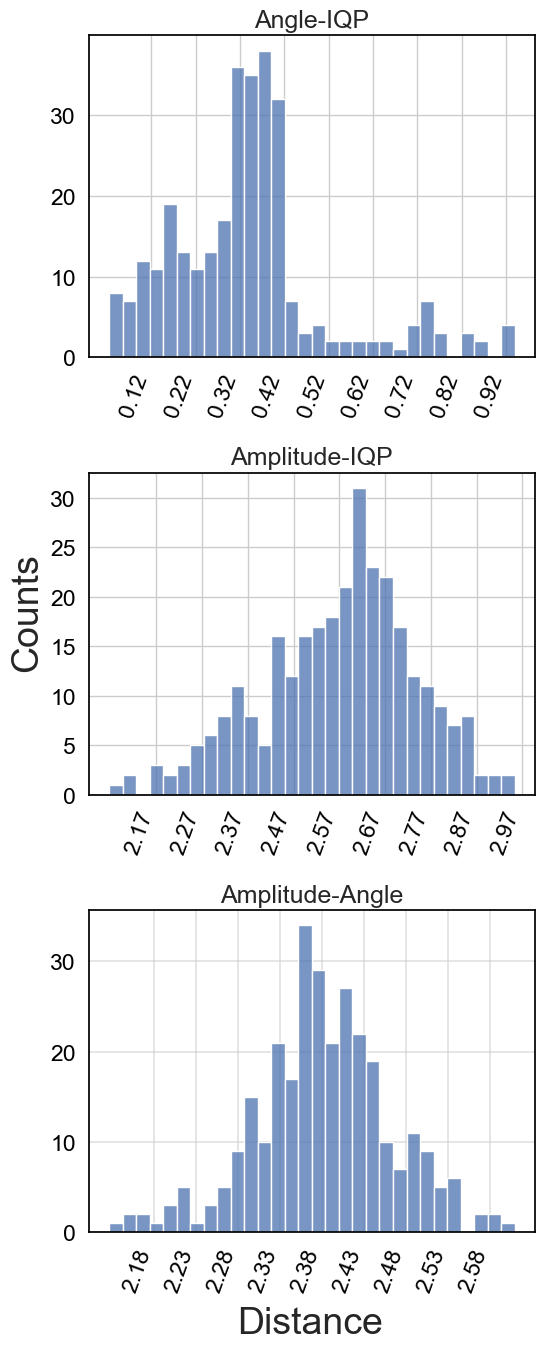}
\caption{To view how the operator pseudo-entropy value for a data point differs between each encoding scheme, for the linearly separable data for dimension $10$, the distribution of distances is displayed. The distances were calculated from Definition \ref{def:diff-metric} and the modulus for the final value.  }
\label{fig:distance-pseudo-entropy-values}
\end{figure} 

It is proposed that the focus of analyzing general encoding schemes should establish a theoretical assurance of full information retention, or at least minimal retention of `important' information. Of course, the concept of information is not well-defined since there are different perspectives of what is implied by information. To demonstrate different perspectives, for a given point-cloud we assume the manifold hypothesis, ergo, the data samples represent an embedded manifold in $\mathbb{R}^n$, call it $M$. Now for a quantum feature map, the map takes points in $M$ to a special unitary operator with a final size of $2^k \times 2^k$. Ergo, the space of special unitary operators, denoted as $ \mathrm{SU}(2^k)$, where $2^k$ is assumed to be the space $\big(\mathbb{C}^2 \big)^{\otimes k}$. Hence, a quantum feature map is a morphism of the form $f:M\to \mathrm{SU}(2^k)$. From here one may stop and analyze questions about retention of topological structure, or a geometric structure that includes geodesic flows, or retention of an algebraic structure, or the retention of information via information theory. 

Observe that the vast majority of techniques--many of the well-used are described and compared in \ref{sec:comparison-other-methods}--consider the expected output of the quantum circuit, adding an extra mapping step beyond the space $f(M) \subset \mathrm{SU}(2^k)$. This extra step is difficult to describe as a flow of information since it is the range of each individual function, i.e., the expected output of a quantum circuit. Hence, the extra step has a large potential of losing information since the functions themselves are not analyzed.

\begin{figure}[th!]
     \centering
    \includegraphics[width=.45\textwidth]{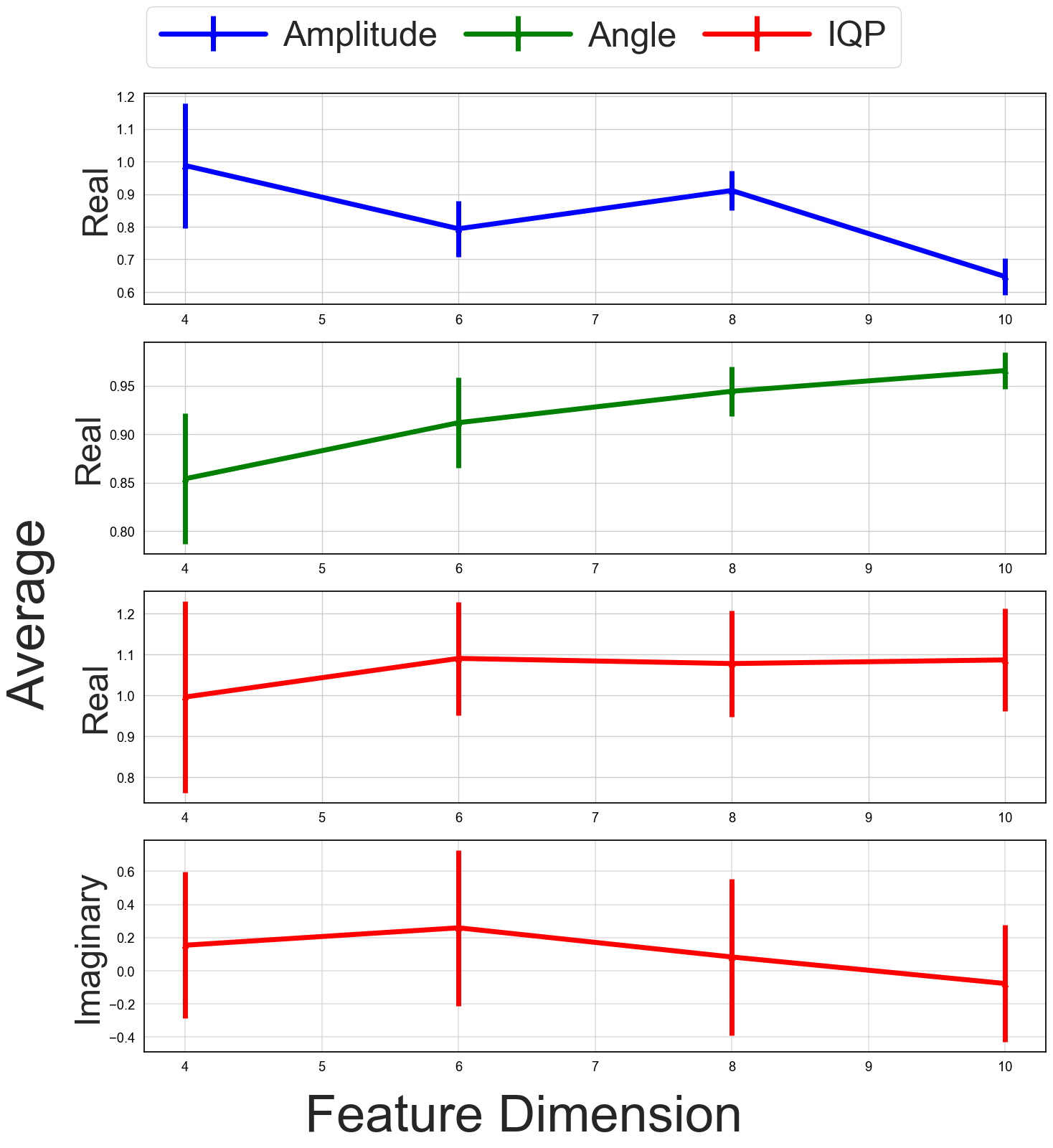}
\caption{Average pseudo-entropy values of linearly separable data over varying dimensions of the features.}
\label{fig:average-pseudo-entropy-values}
\end{figure}

Therefore, a quantum feature map is a morphism where information is either topological \cite{lee2010introduction,gracia2013elements}, geodesic flows \cite{connes1994noncommutative,hatcherbook2003}, an algebraic structure \cite{connes1994noncommutative,hatcherbook2002}, or as an information geometry \cite{amari2016information,NTTTW21}. Information retention is then keeping the structure of a commutative geometry the `same' structure as in the 'pseudo-noncommutative' geometry of $ \mathrm{SU}(2^k)$. The term pseudo is given since the infinite dimensional Lie group of special unitary operators actual holds the spectral triple discussed in \cite{connes1994noncommutative}, but the noncommutativity of finite dimensional is algebraic. Therein lies the difficulty, comparing commutative structures to pseudo-noncommutative structures. However, the areas of commutative, noncommutative, and Lie group geometry are well-studied with a deep and rich literature of research. 

\begin{figure*}[!t]
    \centering
    \renewcommand{\arraystretch}{1.}%
    \begin{tabularx}{\linewidth}{cccc} 
     & Accuracy &  F1 Scores & AUC Scores \\
     \adjustbox{raise = 8 ex}{ \rotatebox{90}{ QSVM } }
    & \includegraphics[height = 3.4cm]{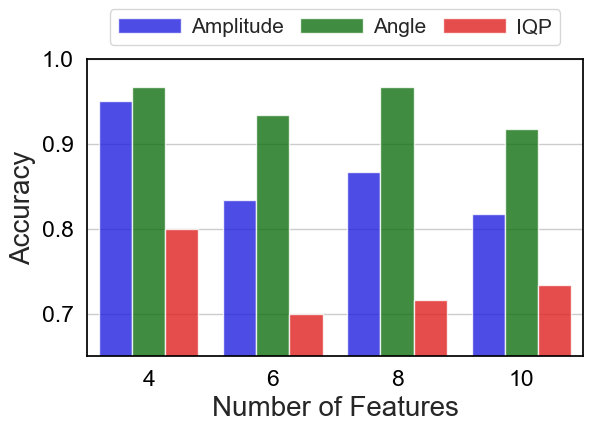} 
    & \includegraphics[height = 3.4cm]{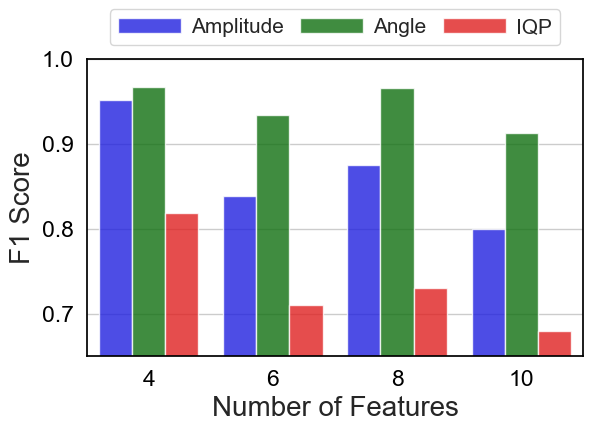}
    & \includegraphics[height = 3.4cm]{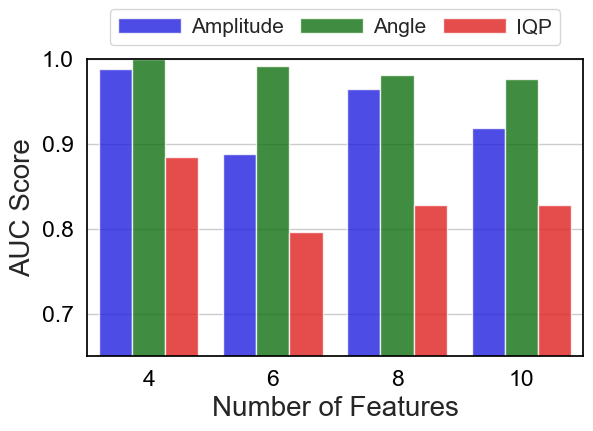}
    \\ \adjustbox{raise = 9 ex}{ \rotatebox{90}{LightGBM} }
    &   \includegraphics[height = 3.7cm]{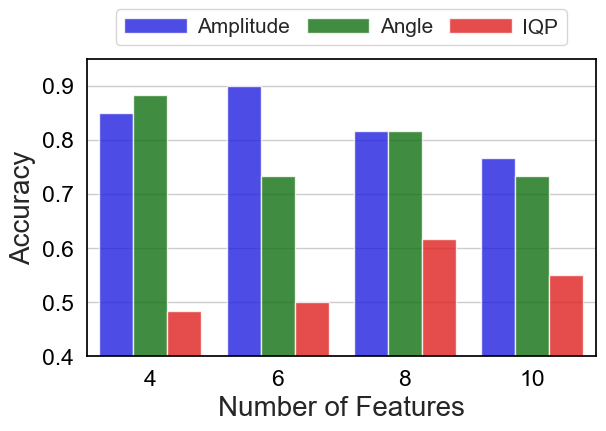}
    &  \includegraphics[height = 3.7cm]{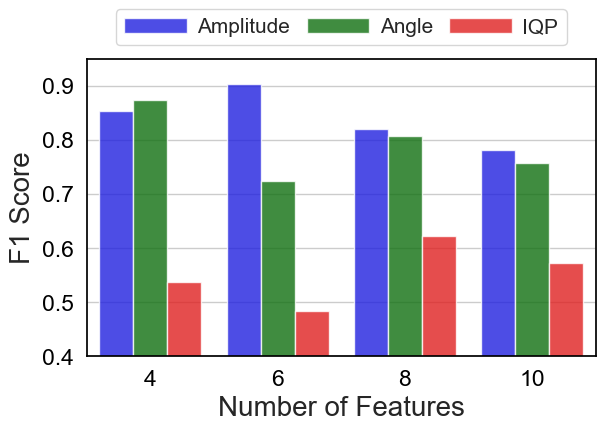}
    & \includegraphics[height = 3.7cm]{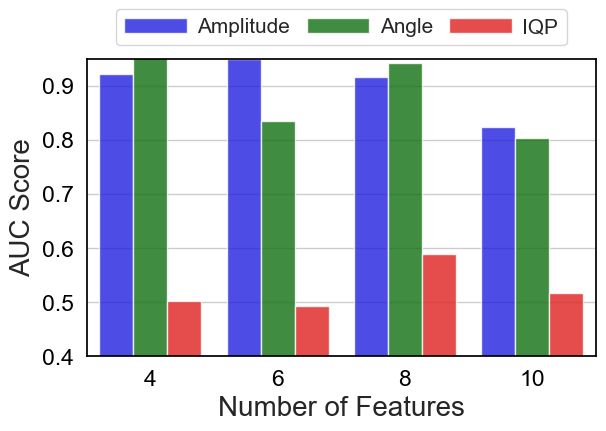} 
    \\
    \end{tabularx}
    \caption{Classification with the QSVM and LightGBM algorithm on different dimensions of the linearly separable.}
    \label{fig:qvsm-LGBM-dim-classify-results}
\end{figure*}

This manuscript focuses on the information geometry of the an embedded smooth manifold and the respective subspace of operators in the space of special unitary operators. 

Before we delve into the derivation of the technique, consider the potential comparison of the entropy of the data and the expected state of a quantum feature map. At first glance, this appears to be the proper perspective, especially since this is the view among many papers on quantum machine learning. With this perspective, the Amplitude encoding scheme would be a perfect fit as each data point is mapped to a point on the standard simplex via a diffeomorphic function. However, Gennaro, Vlasic, and Pham \cite{de2024empirical} empirically displayed that this feature map trained statistical models that respectively underperformed.

\subsection{Experimental Comparison Between Operator Pseudo-Entropy, Entanglement Entropy, and State-Transition Pseudo-Entropy}\label{subsec:experiments}

To empirically display the difference between pseudo-entropy and entanglement entropy, we consider three datasets: linearly separable features, spiral data, and real-world data. For the linearly separable data, we follow the algorithm in Bowles, Shahnawaz, and Schuld  \cite{bowles2024better} to synthesize the data with $10$ features, and for the spiral data we employ the make-moons method in Scikit-Learn \cite{scikit-learn}. Finally, for the real-world data, we consider the Algerian Forest Fires\cite{algerian_forest_fires_547}, in the UCI repository; for simplicity, we only analyzed the data from the Bejaia region, and to prepare the data for the feature maps, the MinMaxScaler from Scikit-Learn \cite{scikit-learn} is applied. Figure \ref{fig:trans-state-pseudo-entropy-data} gives a visual display of these datasets.

For the experiments we considered the encoding schemes of Angle \cite{schuld2019quantum}, Amplitude \cite{grover2002creating,araujo2021divide}, and instantaneous polynomial quantum (IQP) \cite{havlivcek2019supervised}. For the Angle feature map, each feature is mapped to the cube $[0,1]^d$, where d is the number of features, via the sigmoid function. Then each feature is mapped to the single qubit rotation gate, $\displaystyle x \mapsto \bigotimes_{i=1}^{d} \exp\big( -i \arcsin(\sqrt{x_i}) Y_i \big)$, for the Pauli-$Y$ gate on the $i^{th}$ qubit. For the Amplitude scheme, data is mapped to the standard simplex by the diffeomorphism $\displaystyle (x_1,\ldots,x_d,0) \mapsto \frac{ ( e^{x_1}, \ldots, e^{x_d}, e^{0}  ) }{e^{0} + \sum_{i=1}^{d} e^{x_i} }$. Each mapped data point is padded with zeros to ensure the data point is a power of $2$. Finally, for IQP, if all the data points are in the unit cube then each feature is multiplied by $2\pi$, and if not then each respective feature is mapped via the sigmoid function. After the data is prepared data it is then mapped to the operator $\displaystyle x \to H^{\otimes d} \prod_{ i,j = 1 }^{d} \exp\big( -i( \pi - x_i)(\pi - x_j) \cdot Z_i \otimes Z_j \big) \prod_{i=1}^{d} \exp\big(  -i x_i \cdot Z_i  \big)    H^{\otimes d},$ for the Pauli-$Z$ operator on each respective $i^{th}$ qubit.

Figure \ref{fig:pseudo-entropy-example-three-dataset} is a comparison of operator pseudo-entropy, von Neumann entropy and state-transition pseudo-entropy, calculated using basic matrix algebra discussed in the preceding sections. As noted, von Neumann entropy is not granular enough, for instance, the consistent zero value with the Angle encoding scheme. Operator pseudo-entropy and averaged state-transition pseudo-entropy both display a scattered range of values. Interestingly, Figure \ref{fig:pseudo-entropy-example-three-dataset} displays that the operator pseudo-entropy values of the Amplitude feature are all real and positive, although entanglement is essential. This is the scenario that shows it is not necessarily true that entanglement implies complex pseudo-entropy values. Contrarily, all of the pseudo-entropy values for IQP are all complex.

To observe how the operator pseudo-entropy value of a data point differs between an encoding scheme, using only the linearly separable data with dimension $10$, Figure \ref{fig:distance-pseudo-entropy-values} gives the distribution of these distances. The distances were calculated from Definition \ref{def:diff-metric} and then applying the modulus for each distance. The compact range of values for IQP and Angle are further displayed by the small distances between them. Amplitude, on the other hand, has a much larger distance between both IQP and Angle.

Expanding the linearly separable synthetic data into varying dimensions, Figure \ref{fig:average-pseudo-entropy-values} displays the average and standard deviation of the operator pseudo-entropy values. In all encoding schemes, the standard deviation decreases with respect to the dimension of the features. However, the deviation from IQP stays respectively large.   

Further expanding the numerical analysis of the linearly separable data, Figure \ref{fig:qvsm-LGBM-dim-classify-results} shows the efficacy of the binary classification statistical models trained from the quantum support vector machine algorithm, with the package taken from Qiskit \cite{qiskit2024}, and the LightGBM gradient booster algorithm \cite{ke2017lightgbm}. The purely quantum model and hybrid model are both considered to negate potential bias and yield a more general analysis. The training consisted of a 80/20 train-test split, and for each dimension a different seed was applied to the Scikit-Learn \cite{scikit-learn} train-test split method. The LightGBM had a fixed learning rate of $.1$, the number of leaves was $\min\{31, 2^{\mbox{number of features} }\}$, the max depth and number of estimators were set to the number of features, and the evaluation metric is logloss.

All statistical models trained with IQP mapped features are either underperformed or just a 'coin flip'. It needs to be noted that the statistical models trained with the QSVM algorithm performed decently. Contrarily, statistical models trained with either the Angle mapped features or Amplitude mapped features performed well. Interestingly, neither feature map trained models that consistently outperformed the other models. However, models trained with the Angle encoding scheme and QSVM did outperform all other models, including all models trained with LightGBM. 

From the observations, one may hypothesize that when the operator pseudo-entropy are real, or when the state-transition pseudo-entropy values have small values with a relatively small standard deviation, then the quantum feature map will assist in training well-performing statistical models. Particularly, feature maps with real operator pseudo-entropy values ensure data points are not concentrated, allowing for the algorithm to distinguish the classes within the data. In fact, we will go a step further and formalize the hypothesize below, and explore the hypothesis in Section \ref{subsec:svd-dist-values}. 

\begin{hypothesis}\label{hypoth:not-concentrated}
Quantum feature maps that yield small imaginary values generates quantum states that are not exponentially concentrated.
\end{hypothesis}

\subsection{\label{subsec:svd-dist-values}Observation on SVD and Eigenvalues}
Recall that when an operator has symmetric eigenvalues then the pseudo-entropy value will be real and positive. Furthermore, it was shown in Proposition \ref{prop:pseudo-entropy-real-values} that if a circuit has only single qubit operators then the tensor of the operators will have symmetric eigenvalues. From the experiments on the the three datasets of linearly separable, spiral, and Algerian fire, while the Amplitude and IQP encoding schemes have intricate entanglement, Figure \ref{fig:pseudo-entropy-example-three-dataset} showed that the pseudo-entropy value for Amplitude was strictly real for every single data point, while IQP had the majority of values complex. Furthermore, the pseudo-entropy values for Amplitude were not as concentrated.  

\begin{figure}[!ht]
    \centering
    \begin{subfigure}[b]{0.4\textwidth}
    \includegraphics[width=.95\textwidth]{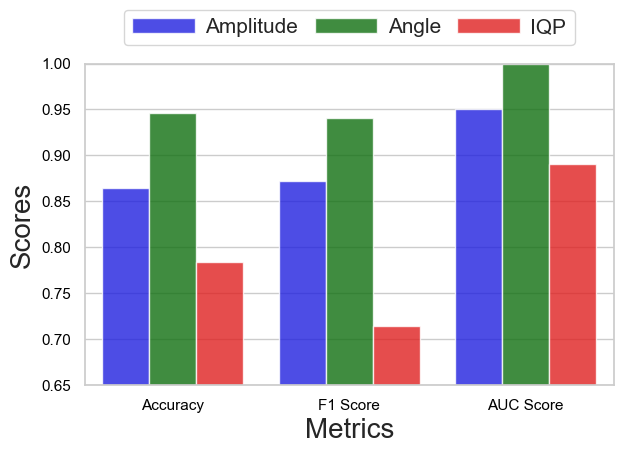}\caption{Classification with QSVM package in Qiskit.\label{fig:alger-qsvm}}
     \end{subfigure}
     \hfill
     \begin{subfigure}[b]{0.4\textwidth}
    \includegraphics[width=.95\textwidth]{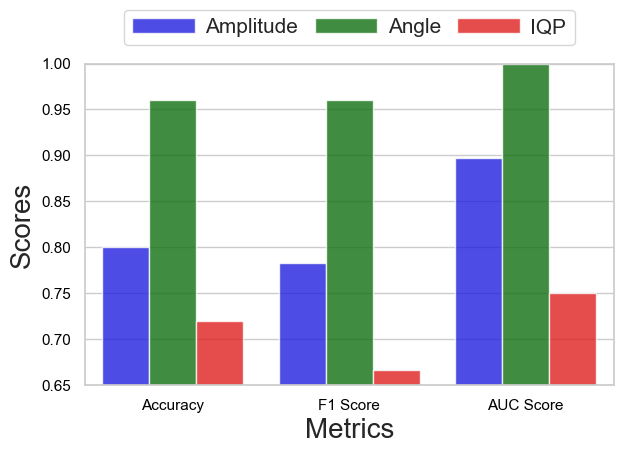}
\caption{Classification with LightGBM.
\label{fig:alg-lightgbm} }
     \end{subfigure}
\caption{Binary classification of the Algerian fire data.}
\label{fig:algerian-classi}
\end{figure}

\begin{figure*}[!t]
    \centering
    \renewcommand{\arraystretch}{1.}%
    \begin{tabularx}{\linewidth}{cccc} 
     & Amplitude & IQP & Angle \\
     \adjustbox{raise = 6 ex}{ \rotatebox{90}{Linearly Separable} } 
    & \includegraphics[height = 5.2cm]{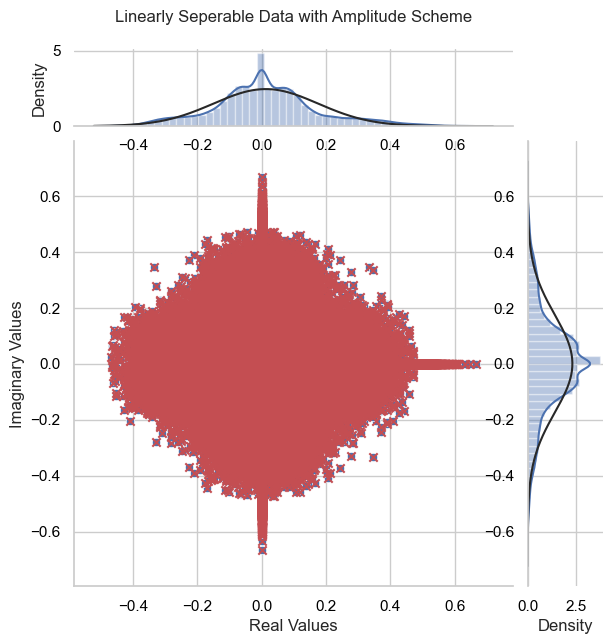} 
    & \includegraphics[height = 5.2cm]{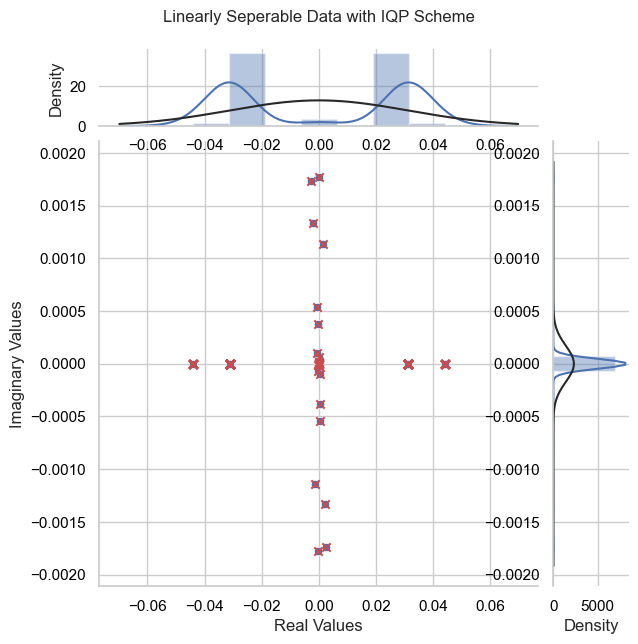}
    & \includegraphics[height = 5.2cm]{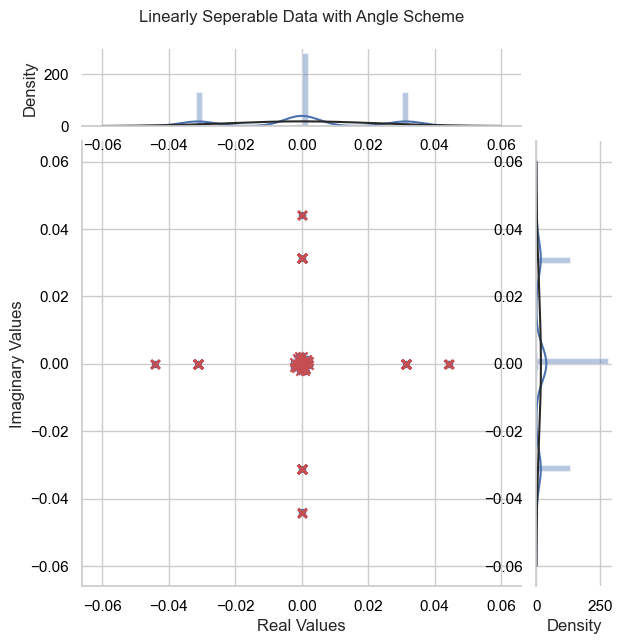}
    \\ 
    \adjustbox{raise = 10 ex}{ \rotatebox{90}{Spiral}}  
    &   \includegraphics[height = 5.2cm]{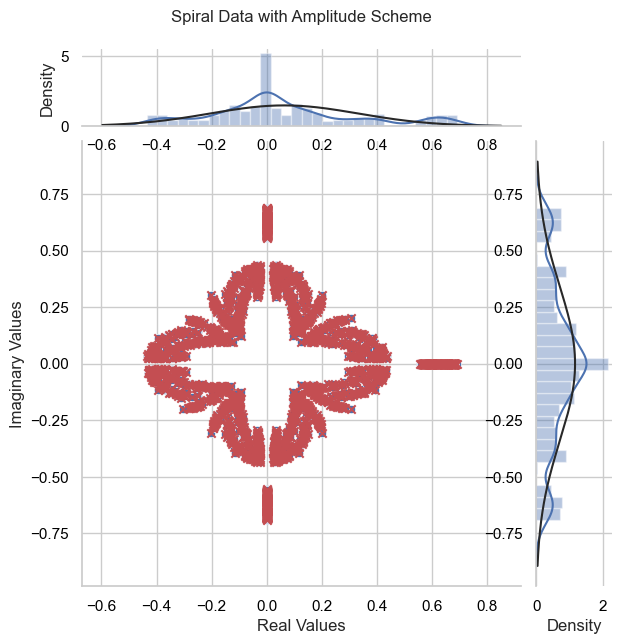}
    &  \includegraphics[height = 5.2cm]{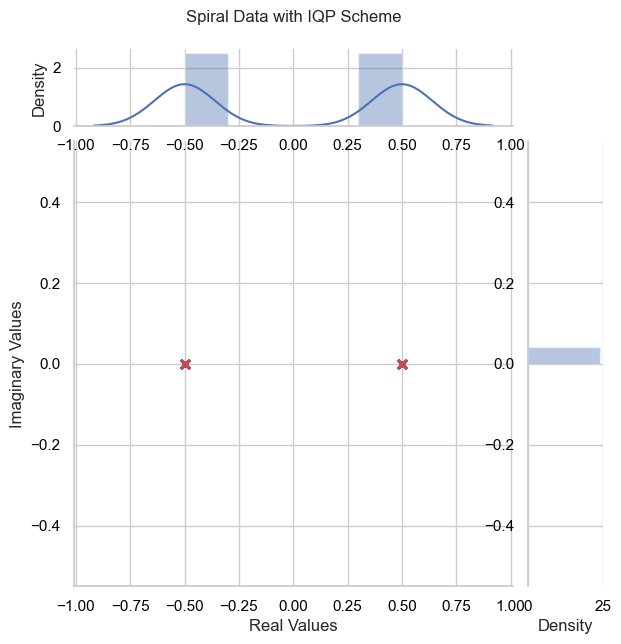}
    & \includegraphics[height = 5.2cm]{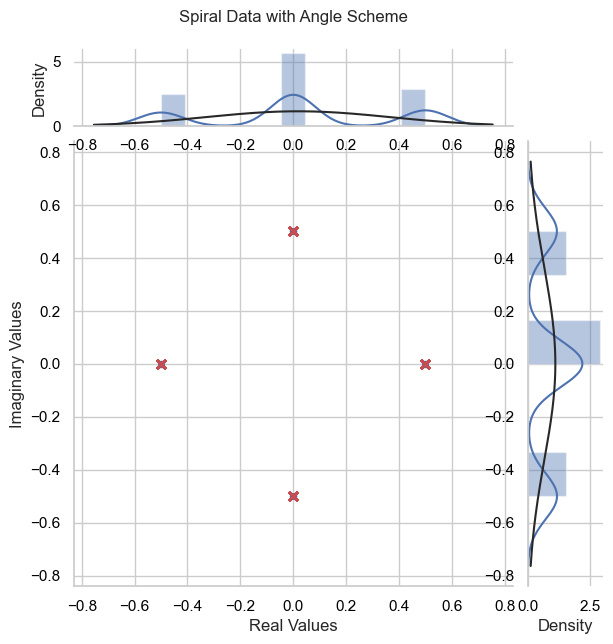} 
    \\ 
    \adjustbox{raise = 7 ex}{ \rotatebox{90}{Algerian Fire} } 
    &  \includegraphics[height = 5.2cm]{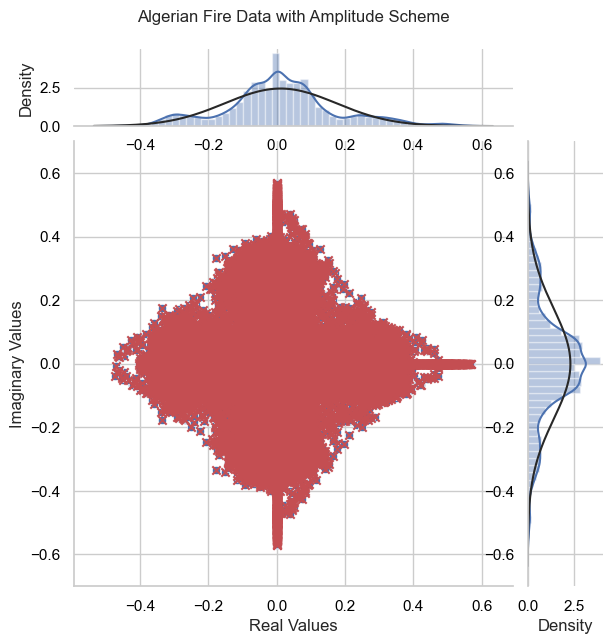} 
    &  \includegraphics[height = 5.2cm]{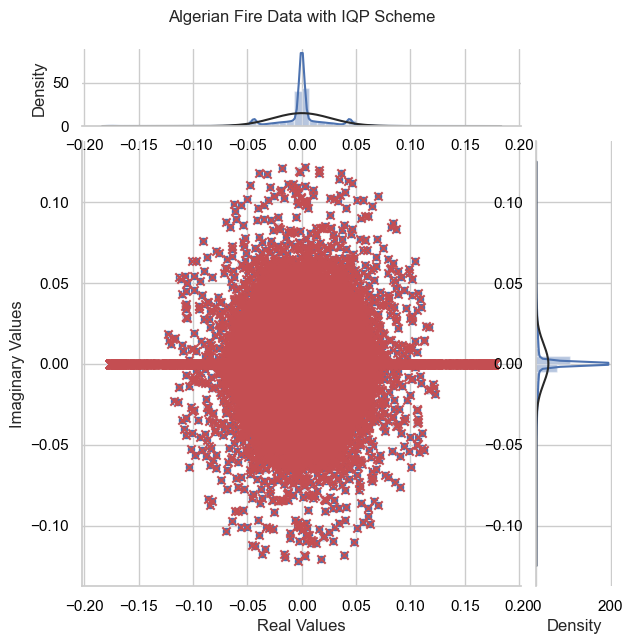} 
    &  \includegraphics[height = 5.2cm]{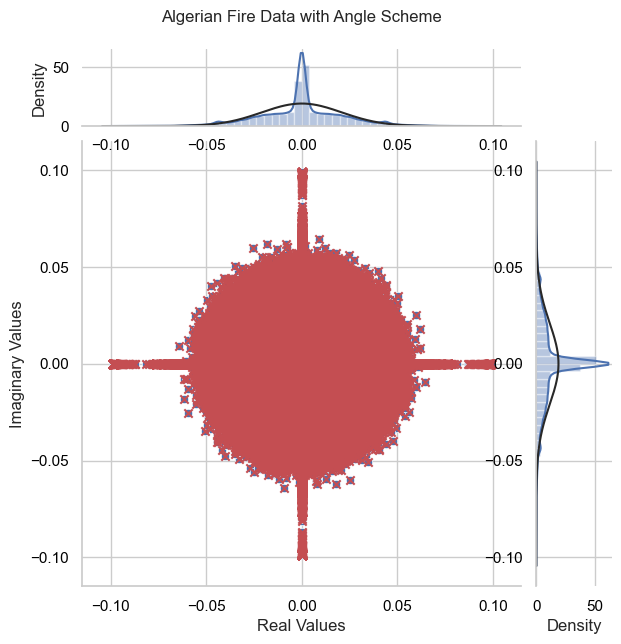} 
    \end{tabularx}
    \caption{ Taking the distribution of unique values from the SVD of the encoding schemes of Amplitude, IQP, and Angle, with every element in the respective datasets of linearly separable, spiral, and Algerian fire. }  
    \label{fig:svd_distribution}
\end{figure*}

Considering the concentration of states noted in Thanasilp et al. \cite{thanasilp2024exponential}, one may ask if the existence of asymmetric eigenvalues, coupled with large imaginary pseudo-entropy values implies a high concentration of states. If such a statement were true then, for a data set $M$ and data points $x$, the SVD of the feature map $U(x)=V_x^{\dagger} \Lambda_x V_x$ would yield a highly concentrated distribution of unique entries in $V_x$, collected for all $x\in M$. This is precisely what is shown in Figure \ref{fig:svd_distribution}. Interestingly, there is a similar characteristic of concentration with the Angle feature map.      

Addressing the entropy generated from the feature maps Amplitude and IQP, first observe the operators that create the entanglement are quite different. The IQP method takes a quadratic operator on two qubits and imposes a pairwise entanglement. Amplitude, on the other hand, takes an increasing and cascading multi-control rotation in such a manner that every 'local' state initiates a control. Hence, the Amplitude encoding scheme ensures, for non-zero entries in the array, are spanned in the state space. IQP is more data driven than state driven. This glaring difference illuminates the results of the numerical experiments, as well as the distribution of values in Figure \ref{fig:svd_distribution}. 

From these observations we may refine and add to Hypothesis \ref{hypoth:not-concentrated} by positing that an encoding scheme that yield large imaginary operator pseudo-entropy values generates a high concentration of entry values in the unitary matrix from SVD.

\section{\label{sec:comparison-other-methods} Comparison of Existing Methods}
We finish the analysis of operator pseudo-entropy by comparing against three well-used techniques to analyze a quantum circuit: expressivity, expressibility, and exploiting symmetry. We demonstrate that operator pseudo-entropy can essentially generalize each technique. To be thorough, mathematical analysis will be required to rigorously demonstrate this generalization, but this is left for future research.   

\subsubsection{Expressivity}\label{subsec:expressivity}

Schuld, Sweke, and Meyer \cite{schuld2021effect} derived a method to empirically test the expressivity of a quantum circuit by calculating the expectation value measurement, taking the partial Fourier series, and determining which complex function represents this expectation. Specifically, the authors analyze the function $\displaystyle f_{\theta}(x) = \bra{0} U^{\dagger}(x,\theta) \mathcal{M} U(x,\theta) \ket{0}$, where $U(x,\theta)$ is a variational operator with $L$-layers and potential repeated uploads of the data and $\mathcal{M}$ is an observable. For simplicity, since $\theta$ is fixed, the term is dropped. Assume the operator has the form $U(x) = W^{(L+1)}S(x)W^{(L)}S(X) \ldots W^{(2)}S(x)W^{(1)}$, where $L$ is number of the layers $L$ The taking $d$ as number of wires in the circuit, we may rewrite the expectation function as $\displaystyle f(x) = \sum_{\mathbf{k}, \mathbf{j} \in [d]^L } e^{i(\Lambda_{\mathbf{k}} - \Lambda_{\mathbf{j}} )x} a_{\mathbf{k}, \mathbf{j} }$. With this notation $[d]^L$ is the set of any $L$ integers between $1,\ldots,d$, $\Lambda_{\mathbf{j}} = \sum_{i=1}^L \lambda_{j_i}$, which is the sum of eigenvalues, and $\displaystyle a_{\mathbf{k}, \mathbf{j} } = \sum_{i,i'}(W^*)^{(1)}_{1 k_1} (W^*)^{(2)}_{j_1 j_2} \ldots (W^*)^{(L+1)}_{J_l i} \mathcal{M}_{i,i'} \times W^{(L+1)}_{i'j_l}\ldots W^{(2)}_{j_2 j_1} W^{(1)}_{j_1 1}$, which just the measurement of the variational layers.  

Since the expectation function $f(x)$ is directly affected by the eigenvalues, which is quite natural and not at all surprising, it appears there is a relationship between expressivity and pseudo-entropy. Of course, since the circuit is measured and the function is the expected value, expressivity has a more intricate form over pseudo-entropy. However, while pseudo-entropy directly measures the energy of an operator, expressivity implicitly measures the energy through function representation. This implicit measuring gives a high score to the Amplitude encoding scheme, yielding a potential for a circuit to create noise.

\begin{figure}[!ht]
    \centering
    \begin{subfigure}[b]{0.4\textwidth}
    \includegraphics[width=.95\textwidth]{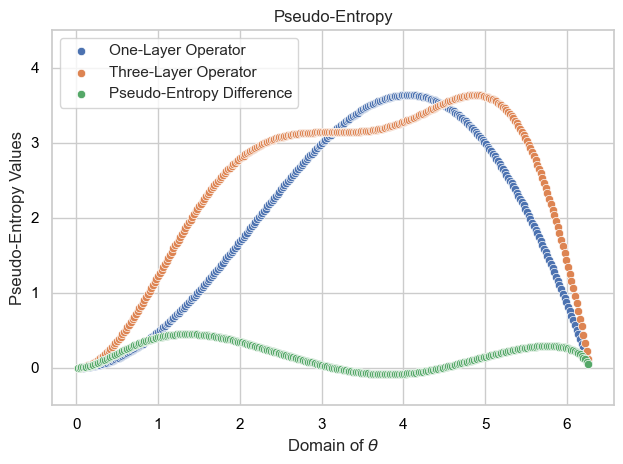}\caption{For the operators one-layer and three-layer, this is the difference of the pseudo-entropy values taken with the metric defined in Definition \ref{def:diff-metric} .\label{fig:pseudo-entropy-essp-example}}
     \end{subfigure}
     \hfill
     \begin{subfigure}[b]{0.4\textwidth}
    \includegraphics[width=.95\textwidth]{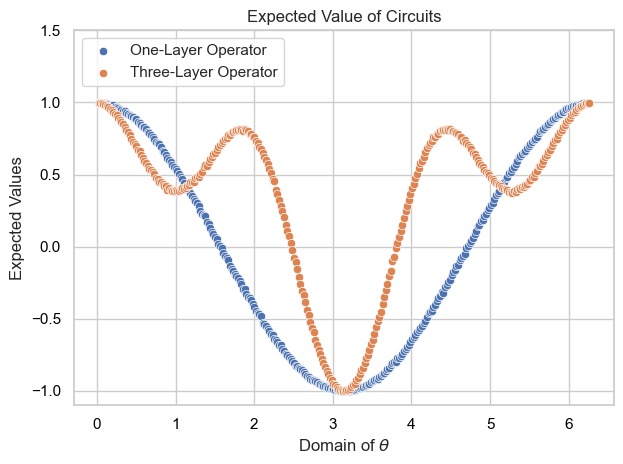}
\caption{The expected values for the circuits of the operators one-layer and three-layer.
\label{fig:expected-essp-example} }
     \end{subfigure}
\caption{Comparisons of operator pseudo-entropy against expressibility and expressivity.}
\label{fig:copmare-method}
\end{figure}

To illuminate the last statement, take the `warm-up application' example in \cite{schuld2021effect}. In this example the authors observe that for a single qubit operator the rotation operator has a simple function representation, and repeated data-upload generates intricacy. For this example we consider the interval $(0,2\pi)$ and the operators $\displaystyle H = \frac{1}{\sqrt{2}} \left[ \begin{array}{cc}
  1   & 1 \\
  1   & -1
\end{array} \right]$, which is the Hadamard gate, and $\displaystyle S = \left[ \begin{array}{cc}
  1   & 0 \\
  0   & i
\end{array} \right]$ and $\displaystyle T =  \left[ \begin{array}{cc}
  1   & 0 \\
  0   & \exp\left( \frac{i\pi}{4} \right)
\end{array} \right]$, which are respective fixed values of the phase gate, where $\varphi \in \{ \pi/2, \pi/4 \}$. Consider the following two operators: $H R_z( \theta ) H$, denoted as one-layer, and $T  R_y( \theta ) T^{\dagger} \cdot S  R_x( \theta )  S^{\dagger} \cdot H  R_z( \theta )  H$, denoted as three-layer. Within the given interval of $(0,2\pi)$ and using the curve fitting in the SciPy package \cite{2020SciPy-NMeth} on the two curves in Figure \ref{fig:expected-essp-example}, after calculating the expectation of each circuit, one may fit of the form $f(x) = a\cdot \cos(\omega \cdot x +d) +c$, and a function of the form  $g(x) = a_c\cdot \cos(\omega_c \cdot x +d_c) + a_s\cdot \sin(\omega_s \cdot x +d_s) + c$, respectively. It is quite interesting how the addition of two more layers resulted in the addition of a sine function.

For operator pseudo-entropy, Figure \ref{fig:pseudo-entropy-essp-example} displays the values, as well as the difference described in Definition \ref{def:diff-metric}. The figure displays smooth curvature in the values for both operators. However, unlike expressivity, there is asymmetry, indicating a more complex relationship of the quantum system dictated by the operator. Particularly, the pseudo-entropy values in the approximate interval of $(\pi,2\pi)$ being larger than the values in the interval $(0,\pi)$. Interestingly, pseudo-entropy values at $\theta = 0, 2\pi$ are zero, while the expectation values are $1$ for both operators, which is just the state $\ket{0}$, ergo, no entropy. Finally, the difference is quite sensitive to small values.

\subsubsection{Symmetric Encoding}\label{subsec:symmetry}

Leveraging the natural symmetry of the data, when symmetry is present available, Meyer et al. \cite{meyer2023exploiting} captured this symmetry with a group representation and crafted a general encoding scheme. With a collection of (semi-)labeled points $\{ (\mathbf{x}_i, y_i) \}_{i}$ and the representation of a symmetric group $V_s\in\mathcal{S}$, where for arbitrary $\mathbf{x}$ and function of labels $y$, $y\big( V_s(\mathbf{x}) \big) = y\big( \mathbf{x} \big)$ for all $s \in \mathcal{S}$, a feature map $U$ and some representation of $\mathcal{S}$, call it $U_s \in \SU(2^n)$, must have the property $U\big( V_s(\mathbf{x}) \big) = U_sU(\mathbf{x})U_s^{\dagger}$ for all $s \in \mathcal{S}$. 

One may see the phase gate feature map does not hold the equality $p\big( V_s(\theta) \big) = U_s P(\theta) U_s^{\dagger}$, for the group representation $U_s$ for all $s \in \mathcal{S}$. Particularly, for a point on the circle, $(x,y)$ the function $\mbox{atan2}(y,x)$ yields the Angle $\theta \in (-\pi,\pi)$, and the negative of argument is a rotation of $\pm \pi$, which implies $e^{i\mbox{atan2}(y,-x)} = e^{i\mbox{atan2}(y,x)}e^{i\pi} = e^{i\mbox{atan2}(-y,x)}$; an analytic formula for $\mbox{atan2}(x,y)$ exists but is quite intricate. Working with just the rotation of negatives, there does not exist a $U \in \SU(2)$ such that $p \big( \mbox{atan2}(y,-x) \big) = U p \big( \mbox{atan2}(y,x) \big) U^{\dagger}$.

Observe the quantum feature maps derived in Meyer et al. \cite{meyer2023exploiting}, using the logic in Equation \ref{eq:von-neumann-der}, we have the equality
\begin{equation}\label{eq:symm-vonN}
    \Tr \left( U\big( V_s(\mathbf{x}) \big) \log\Big( U\big( V_s(\mathbf{x}) \big)  \Big) \right) = \Tr\left( U(\mathbf{x}) \log\big( U(\mathbf{x}) \big) \right),
\end{equation}
hence $S \left( U\big( V_s(\mathbf{x}) \big) \right) = S \big( U(\mathbf{x}) \big) $. 

To summarize the argument, a periodic fit function indicates a strong relationship, and a symmetric encoding scheme with a continuous translation function yields a symmetric fit function. However, as there is potential for periodic fit functions outside of symmetric encoding, this technique generalizes the symmetric technique in Meyer et al. \cite{meyer2023exploiting}. 

To finish this subsection, we address the observation in Meyer et al. \cite{meyer2023exploiting} about the existence of more than one symmetric feature map and the importance of checking the expressibility of the schemes in the `Pitfalls to avoid' subsection in their paper. Ergo, ansatz is not enough to derive a feature map. To see this, consider the unit circle with the Angle encoding scheme with parameter values in $(-\pi,\pi)$. The symmetric rotations and reflections can be captured with  Klein's four-group $\mathbb{Z}_2 \times \mathbb{Z}_2$. Following the example in \cite{meyer2023exploiting}, we take the representation of $\displaystyle V_{(1,0)} = \left[ \begin{array}{cc}
  1   & 0 \\
  0   & 1
\end{array} \right]$, $\displaystyle V_{(0,0)} = \left[ \begin{array}{cc}
  0   & 1 \\
  1   & 0
\end{array} \right]$, $\displaystyle V_{(0,1)} = \left[ \begin{array}{cc}
  -1   & 0 \\
  0   & -1
\end{array} \right]$, and $\displaystyle V_{(1,1)} = \left[ \begin{array}{cc}
  0   & -1 \\
  -1   & 0
\end{array} \right]$. Then for the encoding scheme $U_{G}(x_1,x_2) = G(x_1) \otimes G(x_2)$, where $G\in \{ R_y, R_z \}$, one may calculate that $U_G(x_2,x_1) = \mbox{SWAP} U_G(x_1,x_2) \mbox{SWAP}$ and $U_G(-x_1,-x_2) = \big( X\otimes X \big) U_G(x_1,x_2) \big( X\otimes X \big)$. This begs the question as to which one of the two operators to use. Intuitively, it may not matter which scheme to apply since either operator creates a spin, albeit on their respective 'axis' on the Bloch sphere. Given the shortcoming of expressibility demonstrated in Section \ref{subsec:expressibility}, we will calculate the `goodness' of the fit function for both encoding schemes. The respective Spearman correlation and Xicor correlation \cite{chatterjee2021new} coefficients for $U_{R_z}$ are $(.0020,.9613)$, and the respective correlation coefficients for $U_{R_y}$ are $(.0015,.9613)$, implying there is no significant difference between the feature maps, and thus the ansatz holds.

\subsubsection{Expressibility}\label{subsec:expressibility}

Sim, Johnson, and Aspuru-Guzik \cite{sim2019expressibility} constructed a method to test the efficacy of a quantum feature map, denoted as \define{expressibility}. Expressibility refers to the ability of a circuit to generate the full range of pure states with respect to the Hilbert space the quantum gates. For example, with a single qubit how well does a circuit capture states on the Bloch sphere. Particularly, for a given encoding technique, expressibility applies Kullback–Leibler divergence, or symmetric Kullback–Leibler divergence, to the Haar probability measure of the respective dimension of the data and the density of the states outputted from the feature map. Thus, the score is inversely proportional to the expressibility of the feature map. 

It is claimed that the concept of expressibility can be described through operator pseudo-entropy. To see this, consider feature maps with low expressibility, against feature maps with high expressibility. Low expressibility implies that the distribution of quantum states are concentrated around a few states, with the tail of this distribution very small. This characteristic was seen in operator pseudo-entropy where the values have large imaginary values. Dually, as observed, the operator pseudo-entropy values for feature maps with high expressibility have small imaginary values.

\section{\label{sec:wrap-up}Discussion}

This manuscript described a new and novel technique to analyze a given quantum feature through an extension of von Neumann entropy, denoted at operator pseudo-entropy. The method is shown to be mathematically sound and argued to generalize two well-used techniques in QML, as well as argued to extended the symmetric encoding technique. However, deeper mathematical analysis is required to further support the claims of generality. . 

Furthermore, operator pseudo-entropy is numerically compared against state-transition pseudo-entropy from experiments. From the experiments, operator pseudo-entropy is observed to yield similar information to state-transition pseudo-entropy, while requiring far less computational resources. 

Finally, Section \ref{subsec:svd-dist-values} showed a potential relationship between asymmetric eigenvalues, a concentration of quantum states, and a concentration of entries in the unitary matrices from the SVD of the encoding scheme. From existing techniques, it is not clear how to mathematically prove this relationship, and this appear to be an extremely difficult problem. Furthermore, unexpectedly, the distribution of values in Figure \ref{fig:svd_distribution} display symmetry. It is intuitive to be believe that the distributions should reflect the structure of the data in some manner, but there are many subtleties and Riemannian geometry has a large potential to bring deeper insight.

\subsection{Entropy and Category Theory}
We will show a potential area of research to mathematically analyze the connection of entropy and pseudo-entropy through category theory. We posit that there is a quantum feature map that preserves the entropy of each data point, thus, preserving information.

Pseudo-entropy was motivated by considering how a single data point generates energy in system, and with energy comes entropy. Thus, it is natural to consider that a data point may be described through entropy. Once the concept of entropy of data point has been established, it is then natural to ask the relationship between the pseudo-entropy of an operator and the entropy of a data point. 

Taking $\Delta^n$ as the standard simplex, there exists explicit diffeomorphisms where $\mathbb{R}^n \to \Delta^n$. Thus for an embedded Riemannian manifold $M \subset \mathbb{R}^n$, each data point can be uniquely identified by a finite probability distribution. Thus, there is a $1-1$ mapping of entropy to pseudo-entropy as a flow, and in respect to information theory. Since this is a collection of maps compared against a collection of maps, category theory is a natural mathematical structure to analyze the connection. 

Now, putting emphasis that information in this context concerns information theory, it would be prudent to note that entropy is a forgetful functor, or a map between categories that does not keep information, since it loses geometric and algebraic information, see Bradley \cite{bradley2021entropy}. Assuming that $-1$ is not an eigenvalue, one may show that pseudo-entropy is continuous with respect to the point cloud, noting it is obvious that entropy is continuous with respect to the point cloud. Hence, there is a larger collection of maps between these values via a homotopy. 

This is enough information to derive categories between entropy and pseudo-entropy, as well as a way to connect them. We posit that we have built a foundation for categorical analysis.

As a note, data comes in the form of a point cloud and the manifold is ambiguous. Applications of persistent homology \cite{otter2017roadmap,catanzaro2023harmonic}, and persistent entropy \cite{chintakunta2015entropy,merelli2015topological} to the point-cloud to properly sample the data in order to compare entropy and pseudo-entropy through the approximation of the manifold the point-cloud represents. 

\section{Disclaimer}
About Deloitte: Deloitte refers to one or more of Deloitte Touche Tohmatsu Limited (“DTTL”), its global network of member firms, and their related entities (collectively, the “Deloitte organization”). DTTL (also referred to as “Deloitte Global”) and each of its member firms and related entities are legally separate and independent entities, which cannot obligate or bind each other in respect of third parties. DTTL and each DTTL member firm and related entity is liable only for its own acts and omissions, and not those of each other. DTTL does not provide services to clients. Please see www.deloitte.com/about to learn more.

Deloitte is a leading global provider of audit and assurance, consulting, financial advisory, risk advisory, tax and related services. Our global network of member firms and related entities in more than 150 countries and territories (collectively, the “Deloitte organization”) serves four out of five Fortune Global 500® companies. Learn how Deloitte’s
approximately 460,000 people make an impact that matters at www.deloitte.com. 
This communication contains general information only, and none of Deloitte Touche Tohmatsu Limited (“DTTL”), its global network of member firms or their related entities (collectively, the “Deloitte organization”) is, by means of this communication, rendering professional advice or services. Before making any decision or taking any action that
may affect your finances or your business, you should consult a qualified professional adviser. No representations, warranties or undertakings (express or implied) are given as to the accuracy or completeness of the information in this communication, and none of DTTL, its member firms, related entities, employees or agents shall be liable or
responsible for any loss or damage whatsoever arising directly or indirectly in connection with any person relying on this communication. 
Copyright © 2025 Deloitte Development LLC. All rights reserved.

\bibliographystyle{unsrt}
\bibliography{refs}

\appendix*

\end{document}